\newif\ifarxiv
\arxivtrue

\ifarxiv
    \documentclass[10pt]{article}
    \usepackage{authblk}
    \usepackage[round]{natbib}
    \usepackage{fullpage}
    \usepackage{palatino}
    \usepackage{setspace}
    
\else
\documentclass[sageh]{sagej}
\fi

\usepackage{adjustbox}
\usepackage{amsfonts}
\usepackage{amsmath}
\usepackage{amssymb}
\usepackage{array}
\usepackage{booktabs}
\usepackage{braket}
\usepackage{graphicx}
\usepackage{hyperref}
\usepackage{setspace}
\usepackage{xspace}

\hypersetup{colorlinks=true}
\newcommand\BibTeX{{\rmfamily B\kern-.05em \textsc{i\kern-.025em b}\kern-.08em
T\kern-.1667em\lower.7ex\hbox{E}\kern-.125emX}}

\newcommand{\bb}[1]{\ensuremath{\mathbb{#1}}}

\newif\ifshowrevisions
\showrevisionstrue

\newcommand{\revision}[1]{%
  \ifshowrevisions
    {\color{blue}#1}
  \else
    #1
  \fi
}

\usepackage[textsize=small]{todonotes}
\newif\iffinal
\finaltrue 
\iffinal
\presetkeys{todonotes}{disable}{}
\showrevisionsfalse 
\fi

\begin{document}

\title{Quantum Financial Modeling on Noisy Intermediate-Scale Quantum Hardware: 
Random Walks using Approximate Quantum Counting}

\ifarxiv
\author{Dominic Widdows and Amit Bhattacharyya}
\affil{\normalsize{IonQ, Inc.}}
\affil{\normalsize{\texttt{widdows@ionq.com}}}
\fi

\maketitle

\begin{abstract}
    
Quantum computers are expected to contribute more efficient and accurate ways of modeling
economic processes. Quantum hardware is currently available at a relatively small scale, but
effective algorithms are limited by the number of logic gates that can be used,
before noise from gate inaccuracies tends to dominate results. 
Some theoretical algorithms that have been proposed and studied for years do not perform
well yet on quantum hardware in practice. This encourages the development of suitable alternative
algorithms that play similar roles in limited contexts.

This paper implements this strategy in the case of quantum counting, which is used as a component
for keeping track of position in a quantum walk, which is used as a model for simulating
asset prices over time. We introduce quantum approximate counting circuits that use far fewer
2-qubit entangling gates than traditional quantum counting that relies on binary positional encoding.
The robustness of these circuits to noise is demonstrated.

We compare the results to price change distributions from stock indices, 
and compare the behavior of quantum circuits with and without mid-measurement to trends in the housing market.
The housing data shows that low liquidity brings price volatility, as expected with the quantum models.

\end{abstract}

\section{Motivation: Quantum Finance Implementations in 2023}

Quantum computers are expected to enable more sophisticated and accurate modeling of various financial situations. 
The reasons for the high expectations for quantum finance are in some cases thoroughly worked-out
algorithmically: for example, \citet{egger2020quantumfinance} 
survey applications including option pricing and risk management, 
where Monte Carlo simulation methods are commonly used, 
and explain how quantum algorithms for amplitude estimation offer a potential quadratic speedup (by reducing the number of samples needed for the variance of the probabilistic outcomes to converge). 
As with quantum factoring and search, there are solid reasons for expecting quantum computers to perform
well at large-scale problems that are especially challenging for classical computing methods.

In some cases,
the proposed models are simple and concise enough to be simulated
on classical hardware, and now in the early 2020's their 
behavior can be explored on real quantum computers. However, these models have been small.
\revision{Using just a 2-qubit photonics processor, \citet{qiang2016efficient} were able to model a continuous time 
walk on a connected graph with 4 states.}
\todo{An alternative approach to producing a quantum walk is to use a photonics circuit – could the author comment on the advantages/limitations?}
\cite{stamatopoulos2020option} used 3 superconducting qubits for option pricing.
and \cite{zhu2022generative} used 6 trapped-ion qubits to perform generative modeling for correlated stock prices.

The scale of such experiments has been particularly limited by qubit availability and quantum gate accuracy.
For example, the 3-qubit circuit of 
\cite{stamatopoulos2020option} was optimized down to 18 2-qubit entangling gates and 33 single-qubit gates, but even with this small
circuit, error rates in the results ranged from 62\% raw, to 21\% using Richardson extrapolation for error-correction.
This is not surprising, since the accuracies of the single- and 2-qubit gates were estimated at 99.7\% and 97.8\%
respectively, and $0.997^{33} \times 0.978^{18} \approx 0.587$, so the compounded gate error rate is at least 40\%. 

A safe implementation strategy might be to wait for large-scale fault-tolerant quantum computers to become available,
but this runs the risk of missing opportunities in the meantime. Instead, researchers 
such as \cite{stamatopoulos2020option} and \cite{zhu2022generative} try to use currently-available quantum hardware,
and ask whether implementations can be made robust enough to provide value sooner. In the current NISQ era (Noisy Intermediate-Scale Quantum), the scarce
resources include the number of qubits, and also, as seen above, the number of gates, and especially the number of 2-qubit entangling gates. Circuits are sometimes described in terms of width (number of qubits) and depth
(number of dates, or layers of gates), and both need to be minimized. 

Quantum developers sometimes have many suggested designs to start from: quantum information processing has been explored
as an academic field for some decades, and established literature provides many  circuit recipes \citep{nielsen2002quantumcomputation}. 
A natural strategy is to take such designs, consider their NISQ era limitations, and see if there are
alternatives that provide some of the same functionality using fewer qubits or gates.

This paper develops some new examples of this approach, with the basic example of quantum counting. The central novel contribution
of the paper are the approximate quantum counting circuits, introduced in Section \ref{sec:approx_counting}. The motivation
is that quantum counting is used as a component in the implementation of quantum random walks, which are proposed as a model for stock prices,
and also for {\it beliefs} about the future value of stock prices, for the pricing of stock options.
Beliefs are less exact than prices: it is not very important to make sure that an estimate of \$1,000 comes \$1
after an estimate of \$999 and \$1 before an estimate of \$1,001; but it is important to make sure that these are all treated similarly,
and that doubling any of them gives something in the region of \$2,000. The circuits proposed in this paper demonstrate
such properties, albeit approximately, but much more accurately than is currently possible on quantum computers that use
positional binary representations for numbers that strictly follow the axioms of arithmetic. 

A distinct feature of quantum systems including quantum walks is that they behave differently when they are measured, compared
to when they are left to evolve dynamically. Such behavior has been demonstrated with humans in psychology experiments \citep{kvam2015interference,yearsley2016zeno},
and is an important feature in quantum cognition and economics \citep{busemeyer2012quantummodels,orrell2020quantumeconomics}. 
Section \ref{sec:mid-measure} investigates
the simulated behavior of the approximate counting circuits with mid-measurement, and shows that they exhibit 
desirable behavior (in this case, that more frequent measurement tends to reduce the chances of large changes).
\revision{In a departure from many quantum cognition models, the macro effects of beliefs on transactions and prices cannot be described as the decision of particular cognitive agent, and it may be more appropriate to think of quantum models as
representing the beliefs of whole groups of buyers and sellers, and measurements as decisions observed by the whole market.
This theme is considered as part of introducing quantum walks in finance in Section \ref{sec:randwalk}, and the effects
of measurement in Section \ref{sec:mid-measure}, especially with reference to the housing market.}

To begin with, the next few sections review some of the basic quantum
logic gates and how they are put together into quantum circuits, the use of random walks and quantum walks in finance, and 
how these come together to emphasize the practical quantum counting problem.


\section{Quantum Gates Used In This Paper}

In mathematical terms, the key features that distinguish quantum from classical computers
are superposition and entanglement. This section gives a brief summary of how these properties are 
worked with in quantum circuits. Some familiarity with quantum mechanics, especially Dirac notation, is assumed,
so that $\ket{0}$ and $\ket{1}$ are the basis states for a single qubit whose state is represented in the complex vector
space $\mathbb{C}^2$, a 2-qubit state is represented in the tensor product space $\mathbb{C}^2\otimes \mathbb{C}^2 \cong \mathbb{C}^4$
with basis states $\ket{00}, \ket{01}, \ket{10}$ and $\ket{11}$, 3-qubit states are represented in $\mathbb{C}^{\otimes 3} \cong \mathbb{C}^8$
with basis states $\ket{000}, \ket{001}, \ldots, \ket{111}$, and so on. 
For introductions to how linear algebra is written and
used in quantum mechanics, see \citet[Ch 2]{nielsen2002quantumcomputation}, \citet[Ch 2]{orrell2020quantumeconomics}.
Quantum measurement is probabilistic:
if $\ket{\phi}$ is an eigenvector of a given measurement operator, then
a system in the state $\ket{\psi}$ is observed to be in the state $\ket{\phi}$ with probability given by the square of their scalar
product, $\braket{\phi|\psi}^2$ (the Born rule), and if this outcome is observed, the system is now in the state $\ket{\phi}$.

\begin{figure}
    \centering

\begin{tabular}{ccc}
Pauli-X (NOT)
& 
\raisebox{-0.2cm}{\includegraphics[width=3.5cm]{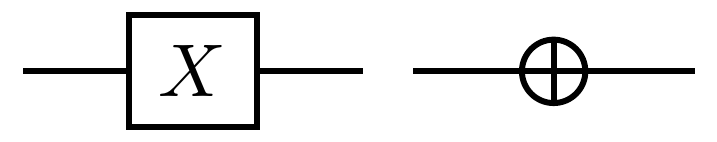}}
&
$\begin{bmatrix} 0 & 1 \\ 1 & 0 \end{bmatrix}$
\\[6 ex]

Hadamard ($H$)
& 
\raisebox{-0.3cm}{\includegraphics[width=2cm]{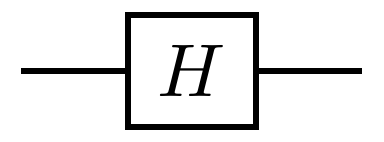}}
&
$\frac{1}{\sqrt{2}}\begin{bmatrix} 1 & 1 \\ 1 & -1 \end{bmatrix}$
\\[6 ex]

\begin{tabular}{c}
$R_X$ rotation \\
$\exp(-i\frac{\theta}{2}X)$
\end{tabular}
& 
\raisebox{-0.3cm}{\includegraphics[width=2.5cm]{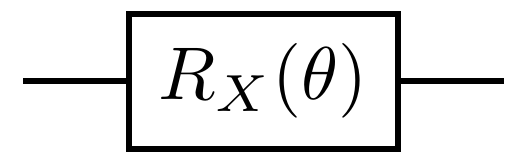}}
&
$\begin{bmatrix} \cos \frac{\theta}{2} & i\sin \frac{\theta}{2} \\ i\sin \frac{\theta}{2} & \cos \frac{\theta}{2} \end{bmatrix}$

\end{tabular}    
    \caption{Some standard single-qubit gates and their corresponding matrices,  
    which operate on the superposition state $\alpha\ket{0} + \beta\ket{1}$ written as a column vector $(\alpha, \beta)^T$.}
    \label{fig:single-qubit-gates}
\end{figure}

Superposition can be realized in a single qubit: the state $\alpha\ket{0} + \beta\ket{1}$ is a superposition
of the states $\ket{0}$ and $\ket{1}$, where $\alpha$ and $\beta$ are complex numbers, with $|\alpha^2| + |\beta^2| = 1$.
Each single-qubit logic is a linear operator that preserves the orthogonality of the basis states and this normalization condition, 
and the group of such operators is $U(2)$, the group of complex $2\times 2$ unitary matrices. Single-qubit gates
that feature prominently in this paper are shown in Figure \ref{fig:single-qubit-gates}. So single-qubit gates coherently manipulate
the superposition state of an individual qubit.

Entanglement is a property that connects different qubits. Since the 1930's, quantum entanglement has gone from 
a hotly-disputed scientific prediction, to a statistical property demonstrated with large ensembles, to a connection
created between pairs of individual particles, to a working component in quantum computers.
All modern quantum computers have some implementation of an entangling gate, and only one kind is really needed,
because all possible 2-qubit entangled states can be constructed mathematically by combining appropriate single-qubit gates
before and after the entangling gate. Furthermore, a single 2-qubit entangling gate and a set of single-qubit gates forms 
a {\it universal gateset} for quantum computing \citep[\S4.5]{nielsen2002quantumcomputation}.

\begin{figure}
    \centering
\begin{tabular}{ccc}
\begin{tabular}{c}
Controlled Not \\ (CNOT, CX)
\end{tabular}
& 
\raisebox{-0.5cm}{\includegraphics[width=2cm]{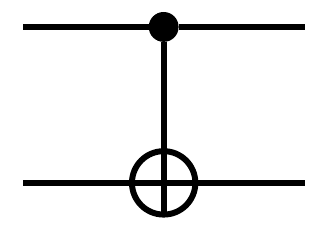}} \quad
&
$\begin{bmatrix}
1 & 0 & 0 & 0 \\
0 & 1 & 0 & 0 \\
0 & 0 & 0 & 1 \\
0 & 0 & 1 & 0 
\end{bmatrix}$
\\[6 ex]

\end{tabular}    
    \caption{The CNOT gate is a 2-qubit entangling gate, that acts upon the
    state $\alpha\ket{00} + \beta\ket{01} + \gamma\ket{10} + \delta{\ket{11}}$. In the standard basis, its behavior can be described
    as ``performing a NOT operation on the target qubit if the control qubit is in state $\ket{1}$''.}
    \label{fig:cnot-gate}
\end{figure}

The CNOT (controlled-NOT) gate of Figure \ref{fig:cnot-gate} is the most common example of a 2-qubit gate in the literature.
In the standard basis, its action is sometimes described as performing a NOT operation on the target qubit if the control qubit
is in the $\ket{1}$ state. Thus, as well as causing entanglement, it is sometimes thought of as a kind of conditional operator in 
quantum programming. Entanglement is the crucial property that distinguishes quantum computing algorithmically, 
because predicting the probability distributions that result from quantum operations with entanglement can become exponentially hard 
for classical computers. In simpler terms, quantum computing is special because it offers special kinds of interference, not because it 
offers special kinds of in-between-ness.

\begin{figure}
    \centering
\includegraphics[width=3cm]{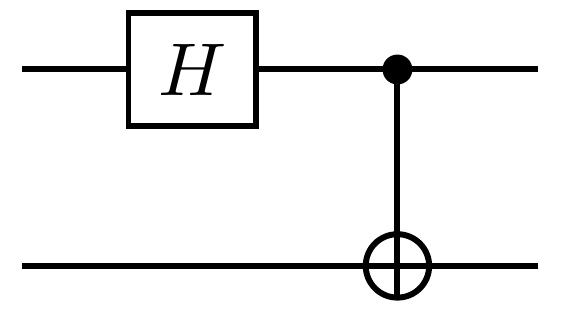}
    \caption{Hadamard and CNOT gates in sequence make a quantum circuit that prepares the Bell state $\frac{1}{\sqrt{2}}(\ket{00} + \ket{11})$.}
    \label{fig:bell-circuit}
\end{figure}

A quantum circuit consists of a register of qubits, and a sequence of logic gates that act on these qubits.
For example, the circuit in Figure \ref{fig:bell-circuit} prepares the famous Bell state (named after physicist John Bell,
whose pioneering theorem motivated experiments that demonstrated real entanglement). It maps the input state $\ket{00}$ to
the state $\frac{1}{\sqrt{2}}(\ket{00} + \ket{11})$, which has the crucial `entangled' behavior whereby if one qubit is
measured to be in the $\ket{0}$ state, the other qubit must also be in the $\ket{0}$ state, and vice versa.

\revision{Quantum circuits finish with measurements that record the 0 or 1 state of at least some of the qubits,
and output this as classical information. (Some platforms also support measuring qubits before the end of a circuit.)
The measurement outcomes are probabilistic, following the Born rule. A run of a single quantum circuit including 
outputting one sample of measurements is typically called a {\it shot}. Circuits are usually repeated several times,
and the output counts from many individual shots are summarized into an output distribution. The process of running all the 
shots and gathering the outcomes is typically called a {\it job}.
}

\begin{figure}
\centering
\begin{tabular}{ccc}
\begin{tabular}{c}
Toffoli \\ (CCX,\\ CCNOT)
\end{tabular}

&
\raisebox{-0.7cm}{\includegraphics[width=1.5cm]{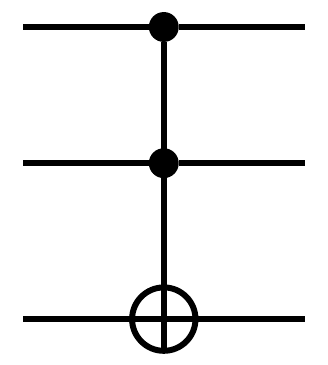}}
&
\raisebox{-0.65cm}{\includegraphics[width=7.5cm]{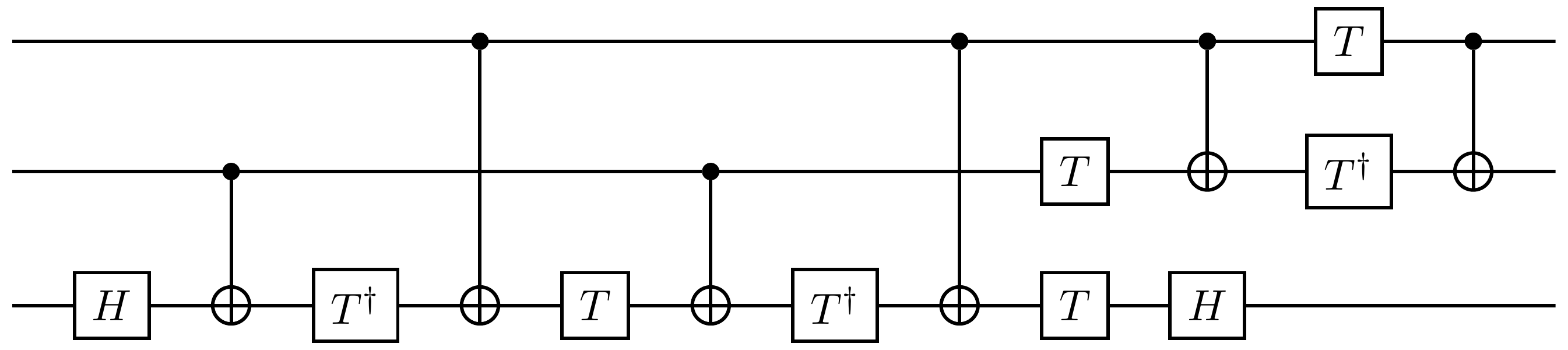}}
\end{tabular}

\caption{The Toffoli gate diagram, showing that it performs a NOT operation on the target qubit if both the control
qubits are in the $\ket{1}$ state. On the right is one of its standard decompositions into CNOT and single-qubit gates, in which 5 CNOT gates are needed
to implement one Toffoli gate.}
\label{fig:toffoli-gate}
\end{figure}

There are many standard gate recipes and equivalences. In particular, larger operators are often thought of as distinct gates
in their own right, an important example being the 3-qubit Toffoli gate of Figure \ref{fig:toffoli-gate}. This is like an extended
CNOT gate --- it has 2 control qubits instead of 1, and performs an X-rotation / NOT operation on the target qubit if both 
the control qubits are in the $\ket{1}$ state. The decomposition in Figure \ref{fig:toffoli-gate} shows that 5 CNOT gates are needed for each Toffoli gate.
There are variants of this, but as a general rule-of-thumb, the error-rate of a Toffoli gate will be at least 4 times the error-rate
of the 2-qubit gates form which it is assembled. Toffoli gates are particularly important for binary arithmetic, as seen in Section \ref{sec:approx_counting}.

In the NISQ era, such considerations are pervasive:
there is a ubiquitous tradeoff between circuit complexity (the number of gates needed to execute a given algorithm) 
and expected circuit accuracy (the more gates we use, the more inaccurate our results become).

\section{Random Walks, Stock Prices, and Quantum Walks}
\label{sec:randwalk}

This section briefly reviews the role of random walks and quantum walks
in the modeling of asset prices. For a more thorough introduction,
see \citet[Ch 7, 8]{orrell2020quantumeconomics}.
A random walk is a mathematical process that constructs a path through some base space
composed of a succession of randomly-chosen steps \citep{xia2019random}. Random walks have been used to model
a range of scenarios including
physical (Brownian) motion, population dynamics, and web browsing sessions,
though they were first proposed for modeling prices of stocks in the Paris Bourse in the work of
 \cite{bachelier1900theorie}.
This was formalized in the Black-Scholes model for pricing financial options: the
paper introducing the Black-Scholes formula assumes that:

\begin{quote}
The stock price follows a random walk in continuous time with a variance rate proportional 
to the square of the stock price. \cite[\S2(b)]{scholes1973pricing}.
\end{quote}

A classical random walk with unit steps up-or-down leads to a binomial distribution, which at large scales is approximated 
by a corresponding normal distribution. Thus, for large simulations, the simplifying assumption of a fixed size for each step
is immaterial, because the overall distribution is normal. However, the most standard formulation for the Black-Scholes 
model assume that the price change for each unit of time is not fixed, but (log-)normally distributed.
The Black-Scholes formula has been widely used as a pricing tool: indeed, over-reliance on
the model, and the amounts of money entrusted to it, have been found to be significant contributors 
to the 2008 market crash \citep{cady2015what,wilmott2017money}. One particular observation is that
the assumption of a constant rate of volatility is not borne out by the long-tail of variations in strike-price, leading
to the claim that a `volatility smile' distribution is a more faithful model in practice 
\citep{orrell2023keep}.

Quantum random walks were introduced in the 1990s \citep{aharonov1993quantumwalks} as a 
counterpart for classical random walks, and
have become a rich and established area of quantum modeling \citep{venegas2012quantumwalks}.
In economics, quantum walks have been proposed as an alternative that takes into account key factors including
varying beliefs or opinions about the future, and the transactions between different traders 
\cite[Ch 7]{orrell2020quantumeconomics}. 
Another anticipated benefit of these quantum walk models is that they will work natively on quantum computers,
when large fault-tolerant quantum hardware is available \citep{orrell2021quantumwalk}.
Quantum walks are thus expected to be a powerful component in pricing models: for example, they may be used to model
the input distributions on which the Monte Carlo methods proposed by \cite{stamatopoulos2020option} depend.

Often the term `quantum walk' is preferred to the term `quantum random walk', 
not only for brevity, but because the internal state of a quantum walk is
typically an entirely deterministic superposition of different states. 
For example, a walk that starts in position 0 with a 50-50 chance of going in either direction will, after one step, be in a superposition of the states representing
positions $-1$ and $+1$, with equal amplitudes in the superposition. 
The quantum state vector representing this superposition can be predicted exactly:
it is only the
measurement outcome that is probabilistic, when one of these distinct possibilities 
is randomly selected.

In the most standard presentation, a quantum walk consists of 
a {\it quantum walker} and {\it quantum coin}. At each turn,
the coin is tossed, and the walker's position moves depending on the 
coin's resulting state. A canonical example is an
unrestricted discrete walk, where the positions correspond to
integers, and each move is a single step, represented by
incrementing the position integer by $\pm 1$.

This leads to an elegant expression for the {\it shift} or {\it translation} operator \cite[Eq. 9]{venegas2012quantumwalks} \cite[\S7.1]{orrell2020quantumeconomics}:
\begin{equation}
  \ket{0}_c \bra{0}\otimes \sum_i \ket{i+1}_p\bra{i}
+ \ket{1}_c\bra{1}\otimes \sum_i \ket{i-1} _p\bra{i}.
\label{eqn:shift_operator}  
\end{equation}

\noindent
The $c$ and $p$ subscripts refer to the coin and position registers.
The positions are represented by integer states $\ket{n}$ for $n\in \bb{Z}$. 

\todo{It should be clarified that the quantum walk model can be used in two different but related ways ...}

\todo{I think there is an onus to argue more clearly about 
possible advantages of quantum random walk relative to classical or at least explain or illustrate more the differences between 
the two}

\revision{The quantum walk model can be used in two different but related ways. Firstly, it can be used to 
model evolving beliefs or opinions, for example, subjective estimates of what a particular asset will be 
worth at a given future time. Different and even contradictory beliefs can be modeled in the quantum walk 
as a coherent superposition of different possible values. Secondly, it can be used to model actual asset prices,
as observed in recorded transactions. A transaction behaves like a measurement on the quantum walk state, 
which forces the coherent superposition into a particular state. Some ways of inserting and tuning the level of 
decoherence are discussed by \citet{orrell2021quantumwalk}, noting in particular that if the quantum walk is measured
at every time increment, decoherence is complete and the quantum walk collapses to the classical version.
}

\todo{\tiny
Pg 8, line 8: “Experiments in simulating quantum walks and harmonic oscillators on real quantum computers have been very small so far.” It would be useful if the author could elaborate on the relationship between quantum walks and the quantum oscillator.
}

\revision{An interesting feature of long, coherent quantum walks is that the distributions become two-tailed, 
as do those of the quantum harmonic oscillator at higher energy levels \cite[Ch 10]{orrell2020quantumeconomics}.
Quantum dynamic models have been used to represent several cognitive scenarios \cite[Ch 9]{busemeyer2012quantummodels},  
and oscillator models in particular have been used to model fluctuations in the stock market \citep{orrell2022oscillator}.
It is intriguing that the distributions produced by quantum walks with and without decoherence are similar to those
produced by oscillator models at high and low energy levels.
}

\revision{This has provided considerable motivation for implementing quantum walk and oscillator models.}
However, experiments in simulating quantum walks and harmonic oscillators on real quantum computers have
been very small so far, restricted to just 2 or 3 qubits, and have reported very noisy results
using superconducting hardware \citep{qiang2016efficient,kadian2021quantumwalk,puengtambol2021implementation}.
The reasons for this are explained in the next section.

\section{Quantum Counting and the Challenge of Recording Position}
\label{sec:counting-challenge}

To simulate a quantum walk using repeated applications of the shift operator in Equation \ref{eqn:shift_operator}, we need to model tossing a coin, and tracking position.
The coin-toss is easy for today's quantum computers to implement
effectively. For example, we use a single qubit and a Hadamard 
gate which acts like a `beam-splitter', putting the coin into a superposition of $\ket{0}$
and $\ket{1}$ states. 

The bigger challenge for quantum computers today is tracking the position:
in other words, the quantum counting problem \cite[Ch 4]{haven2017quantumsocial}. 
For non-negative numbers, the states $\ket{n}$ can be associated with the energy levels in 
a harmonic oscillator \citep{jain2021quantumoscillator}. In theory this might connect the process
of quantum counting with the use of motional modes in for quantum information processing,
but this is not yet available on commercial quantum computers \citep{chen2021vibrationalmodes}.

The most traditional way to represent numbers on a computer is to use some form
of binary positional notation. For example, the binary expression 110 represents the number 6 (or the number 3, if the bits are read in reverse order).
Quantum binary `adder circuits' were designed by Feynman in the early papers that first motivated 
quantum mechanical computing \citep{feynman1986quantummechanical}, but the ongoing presence of errors in NISQ-era machines limits the number
of steps we can reliably count \citep{orts2020review}.

\begin{figure}
    \centering
\includegraphics[width=7cm]{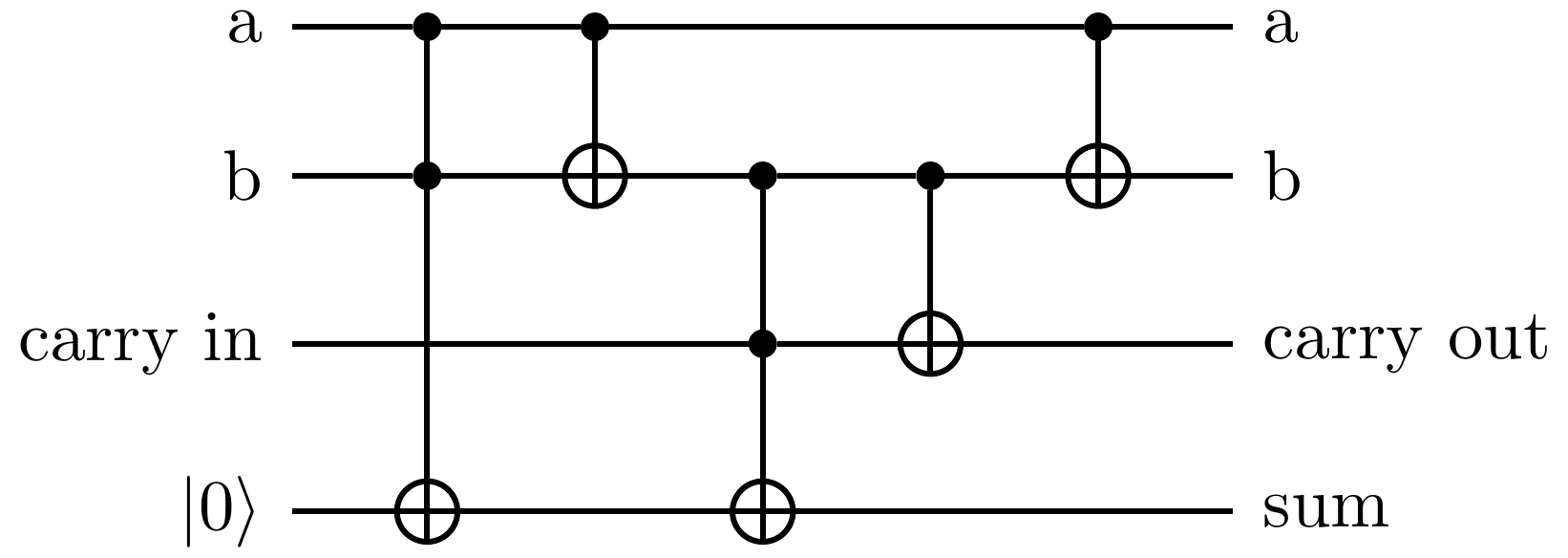}
\caption{A quantum full adder circuit, first introduced
by \cite{feynman1986quantummechanical}, uses
2 Toffoli and 3 CNOT gates, which is at least 11 2-qubit gates.}
\label{fig:quantum_full_adder}
\end{figure}

Choosing a binary positional encoding, as used in classical computing, makes quantum counting very susceptible to 2-qubit gate errors, because manipulating
binary encodings takes a lot of entanglement and coordination between qubits. 
To compute the sum of two binary numbers $A$ and $B$
of bitlength $n$ using classical Boolean algebra, we add (XOR) the least significant bits, and
then at each other position we add the corresponding bits along with
a `carry' from the previous stage, setting the output and passing a `carry' on to the next position. 
 \cite{feynman1986quantummechanical} explained the quantum gate operations needed for each such step
in the quantum full adder circuit of Figure \ref{fig:quantum_full_adder},
and modern versions are optimized variants on this theme \citep{orts2020review}. If it takes 11 2-qubit gates for each
full-adder, then adding two single-byte (8-bit) numbers 
using this approach uses $\sim 80$ 2-qubit gates, so by the time such a register has successively added 10 numbers, the chances of an error are over 50\% even with a two-qubit gate fidelity of 99.9\%, which is on
the high-end of performance estimates at the time of writing \citep{ionq2022aria}.

\begin{figure}
    \centering
    \includegraphics[width=\linewidth]{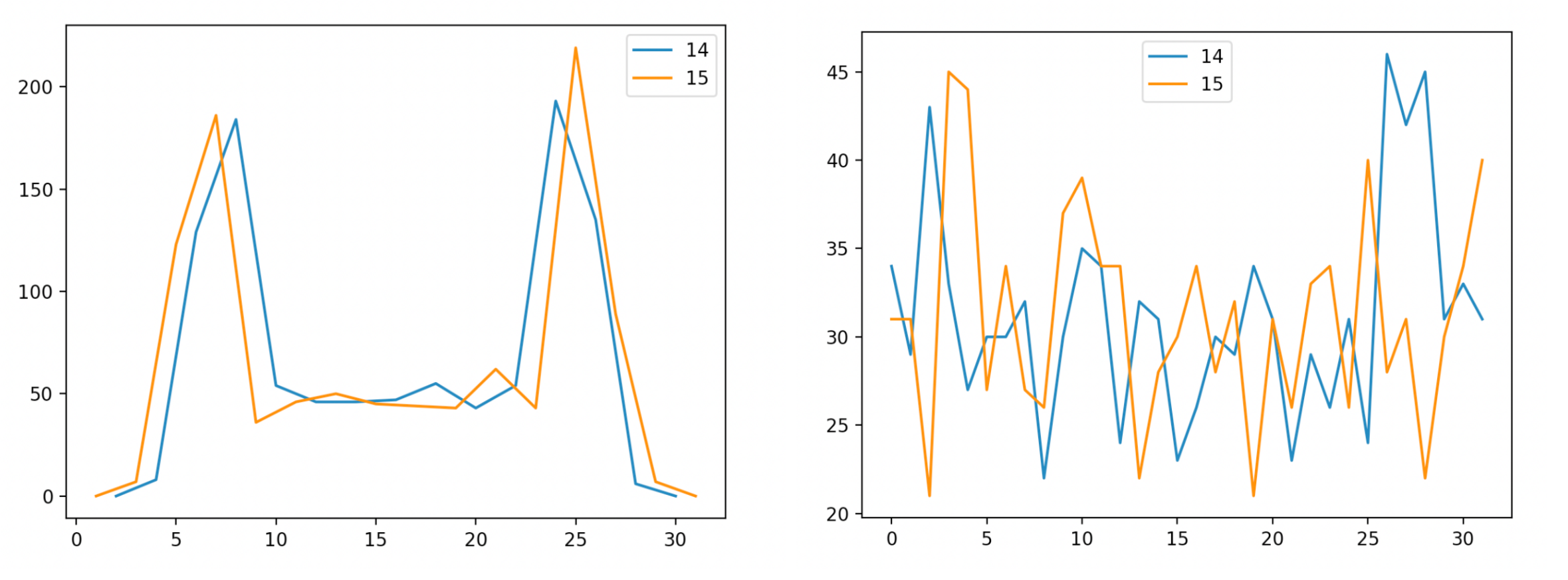}
    \caption{Ideal simulated quantum walk after 14 and 15 steps (left),
    compared with results simulated with expected noise (right). The ideal distribution
    has the two-tailed peaks characteristic of a quantum walk or harmonic oscillator, but 
    this quickly gets lost with noise (right).}
    \label{fig:quantum-noisy-walk}
\end{figure}

Error rates with quantum counting can thus undermine the simulation
of quantum walks, and block this application of quantum finance. 
The problem is demonstrated in Figure \ref{fig:quantum-noisy-walk}, which
compares quantum walks with ideal outcomes and with noise. This shows the 
vulnerability of the quantum counting process to noise after a handful of steps.
\revision{It should be noted that {\it some} noise or decoherence can be useful in 
quantum models, for example, in deriving price estimates from many quantum walk simulations 
\cite[Ch 6]{orrell2020quantumeconomics}. The predictions later in this 
paper are also averages over many somewhat-noisy circuits: the key engineering
challenge here is to understand the noisy behaviors, and find those that work
well enough to enable the task at hand.}

\todo{Pg 8, line 43: The paper states that “Error rates with quantum counting can thus undermine the simulation of quantum walks, and block this application of quantum finance.” Also the caption for Figure 6 notes that the bimodal property “quickly gets lost with noise”. However it should be noted again that there are cases where a higher degree of noise is desired (as when modelling asset price distributions).}

There are many optimizations and alternatives. 
We expect progress in quantum hardware to enable greater fidelity 
and stability, and eventually mid-circuit quantum error-correction
should make the current problems with quantum counting obsolete --- but this comes at the cost of waiting for fault-tolerant quantum computing.
Quantum addition algorithms can be optimized \citep{cuccaro2004new,gidney2018halving}, and the incremental operation
of counting can be made simpler than repeated full-register addition
\citep{li2014class}. An interesting benchmark challenge
could be to design and evaluate quantum circuits and see how far they can
count with $>50\%$ fidelity, but that task is not undertaken here,
because none of these methods simplify counting
enough for many successive counting operations on nontrivial quantum registers
to be performed accurately yet.
Instead, this paper proposes alternative circuits that can be
used to simulate steps and positions in a walk, without requiring
exact counting.

\section{Approximate Quantum Counting: Fault-Tolerant Circuits for Tracking Position}
\label{sec:approx_counting}

By now, the central modeling problem of this paper should be clear: we would like
to be able to model a quantum walk on a quantum computer, but the use of positional 
binary notation to represent integer quantities requires `increment' and `decrement' operators that require too many entangling gates to give accurate results on current quantum hardware. 

To avoid these pitfalls, we introduce alternative circuit designs that can also be used for recording position in a random walk. 
Instead of trying to ensure that every move goes up and down by exactly one
step on the position axis, the position register is incremented using some gate combination
that is likely to move the position by some amount that is generally positive for
upward steps, and downwards for downward steps. Another way to describe this is that
instead of putting all the randomness in the coin toss and following this with deterministic
shift operators, we toss a random coin and then combine this with a somewhat-random shift operator.
For larger circuits, such methods can give a walk that goes up and down more reliably than the results
we get if we try to insist that the position represented by the state $\ket{n}$ is an accumulation of exactly $n$ steps of unit length in that direction.

\revision{
The reader should note that all the results from the circuits in this section are statistical in nature, so
the results are naturally noisy in various ways. Some of the randomness is purely quantum, in that different runs of the 
same circuit are expect to give different measurements according to the Born rule, and running jobs with more shots
gives closer statistical approximations to the `real' quantum distribution. For some of the design patterns in this section, randomness is also due to the way the circuits are constructed, so this randomness is expected and is classical in nature.
Also, NISQ-era results also include noise from hardware errors. It is not always immediately obvious which sources
of noise or randomness are most impactful for a given design, though some of the key features are noted.
}

\subsection{Arc Counter Circuits}

The Arc Counter circuit design uses only single-qubit rotations throughout. Such a representation
is sometimes called a rotation encoding \citep{schuld2021machine}. It is particularly
easy to implement for a modest number of features, and can be incremented as new feature weights are encountered.

\begin{figure}
    \centering
    \includegraphics[width=\linewidth]{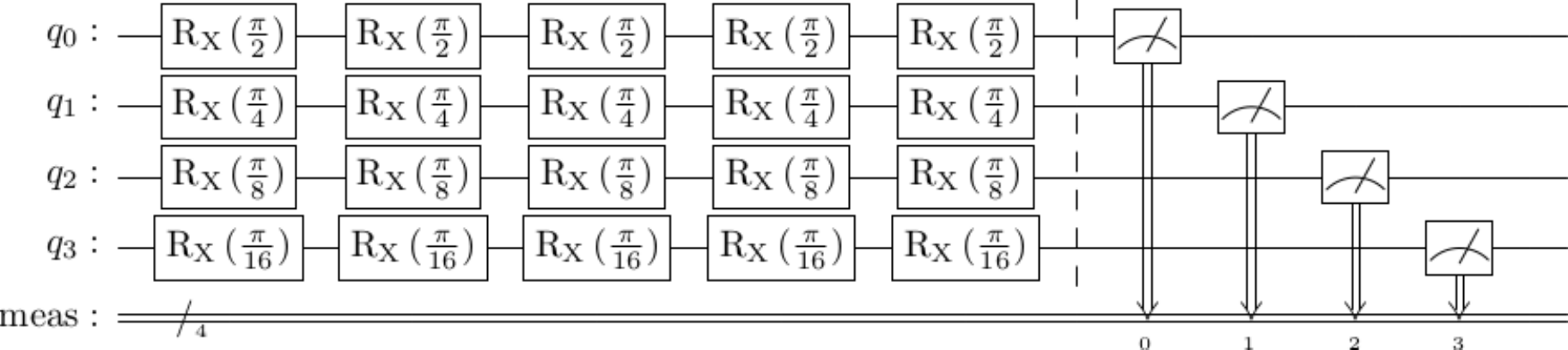}
    \caption{Arc counter circuit}
    \label{fig:arc_counter_circuit}
\end{figure}

To use a rotation encoding for counting, qubits are rotated through particular arc-lengths or angles at each incremental step.
Each qubit is used to represent one digit in a binary register, where each bit is twice as significant as its predecessor. At each step, each qubit traverses an arc that is inversely proportional to that qubit’s significance, as in Figure \ref{fig:arc_counter_circuit}.
This means that the $n+1^{th}$ qubit rotates at half the rate of the $n^{th}$ qubit. 

A good analogy for this representation is to think of each qubit as one of the hands on a traditional analog clock. On an analog clock, the minute hand cycles at 12 times the speed of the hour hand, and the second hand 60 times faster still, whereas in our binary clock, 
the ratio between the speed of rotation of each successive pair of `hands' is 2:1.
This analogy works well for the standard binary positional notation for integers, which can be
thought of as a binary digital clock. In a digital register (or an abacus), the digits logically depend on one another for correct incrementing, because we need to know that one register is full before we increment the next. By contrast, the hour hand on a clock does not `carry' information
when the minute hand completes a cycle --- it just rotates at its own slower pace. Thus the rotation encoding clock-based design requires much less coordination (and hence entanglement) between the qubits.

This comes at a representational cost --- the register does not represent exact integers, and  
random variations in the outputs are expected, because many fractional angles are used throughout the circuit. (This is true for the basic counting operation, irrespective of whether the counting
is coupled with a `coin toss' operation.)

\begin{figure}
    \centering
    \includegraphics[width=\linewidth]{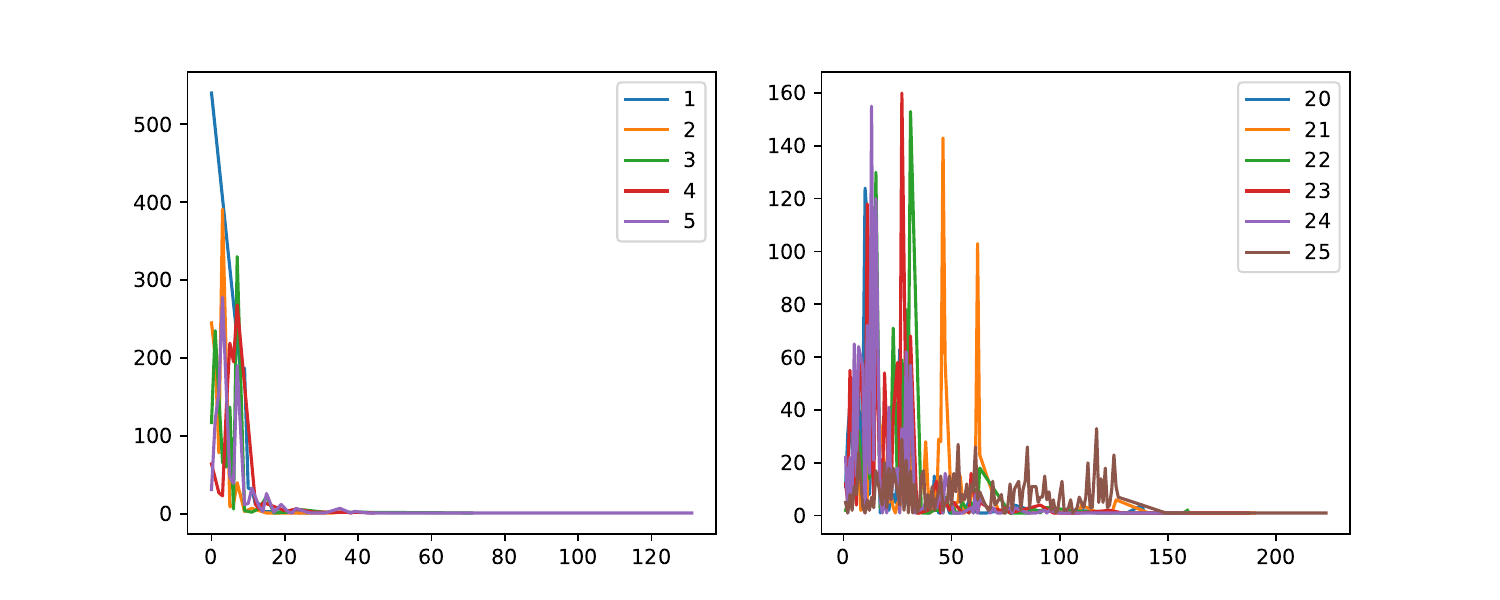}
    \caption{Quantum walk results after different numbers of steps with arc counter circuit
    and an 8-qubit register. 1000 shots for each number of steps.}
    \label{fig:arc_walk_results}
\end{figure}

Sample results from a quantum walk with an 8 qubit register are shown in Figure \ref{fig:arc_walk_results}.
Statistically noteworthy properties include:
\begin{itemize}
    \item The mean distance from the  starting point generally increases with the  number of steps. 

\item In some cases, the position appears to jump ahead, because a high-order qubit is measured in the 
$\ket{1}$ state. 
This can happen (with low probability) after just a single step.

\item There are sometimes peaks in the distributions after specific powers of two or their combinations (e.g., peaks at 48, 64, 96).
It may be possible and desirable to find ways to smooth out these peaks.
\end{itemize}

Since there are no 2-qubit entangling gates, error rates are lower, but there’s also no physical 
quantum advantage from this design --- it is easy to model this distribution on a classical computer. It’s 
possible that such distributions might be useful models for random processes, but this would not require 
quantum computers to simulate. 

\subsection{Reversal and Superposition by Classical Post-Processing}

An important feature of the traditional `binary adder' circuit components is that they are able to decrement (subtract) as well as increment (add). The arc counter circuit, and the others below, do not support this feature. 
The logical work to guarantee that a change from $01111111$ to $1000000$ happens in-concert for all of the bits is precisely what we’ve given up, which makes it much harder to orchestrate a difference between positive and negative steps with large distances.

\begin{figure}
    \centering
    \includegraphics[width=0.8\linewidth]{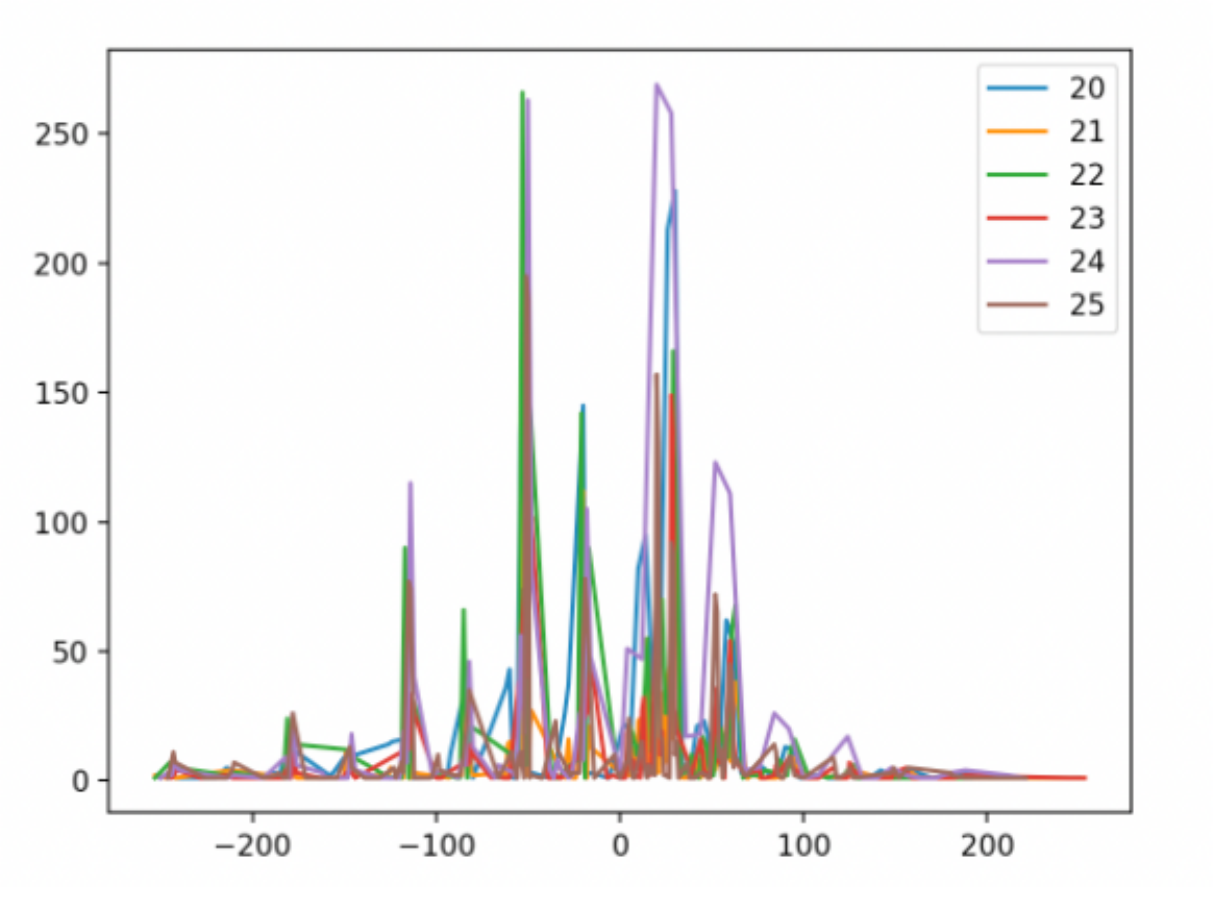}
    \caption{Two-directional arc walk circuit from combining positive and negative distributions. 1000 shots for each number of steps.}
    \label{fig:arc_walks_2_way}
\end{figure}

As noted by \citet[Ch 4]{haven2017quantumsocial}, it is natural for quantum systems to have a lowest state 
which we may call $\ket{0}$, and if we want to generate a full set of integers including the negative
ones, these can be constructed as differences between positive natural numbers. This leads to 
an alternative method for simulating random walks that can evolve in both directions. We use two quantum circuits, one for the `up' steps and one for the `down' steps, and subtract the results from the down circuit from the results from the up circuit as a classical post-processing step. The `up' and `down' circuits can even be configured with different `clock speeds', which has been done in the example of Figure \ref{fig:arc_walks_2_way}.

\subsection{Arc Walk Counter Circuit}

\begin{figure}
    \centering
    \includegraphics[width=\linewidth]{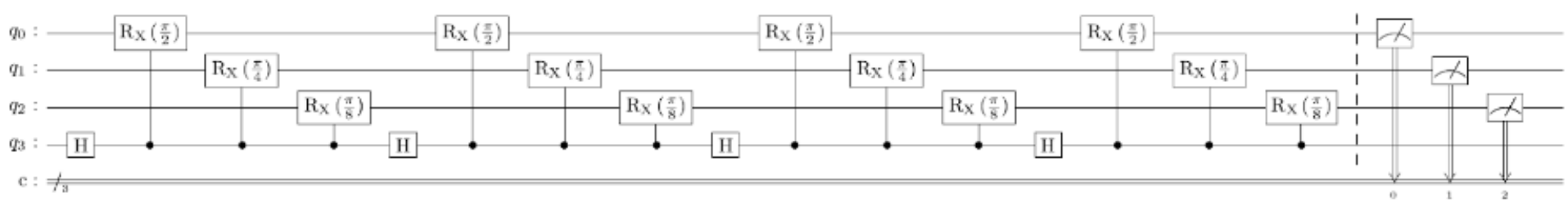}
    \caption{Arc walk counter circuit}
    \label{fig:arc_walk_counter_circuit}
\end{figure}

This circuit design combines the arc rotations of the Arc Counter design above, with the `Hadamard quantum coin' prevalent in the quantum walk literature.
At each step, a Hadamard gate `tosses a coin', and if the outcome is `heads', or $\ket{1}$, a controlled rotation is performed on each of the other qubits, following the same pattern of angles as in the Arc Counter circuit above. This gives the circuit pattern of Figure~\ref{fig:arc_walk_counter_circuit}.

\begin{figure}
    \centering
    \includegraphics[width=\linewidth]{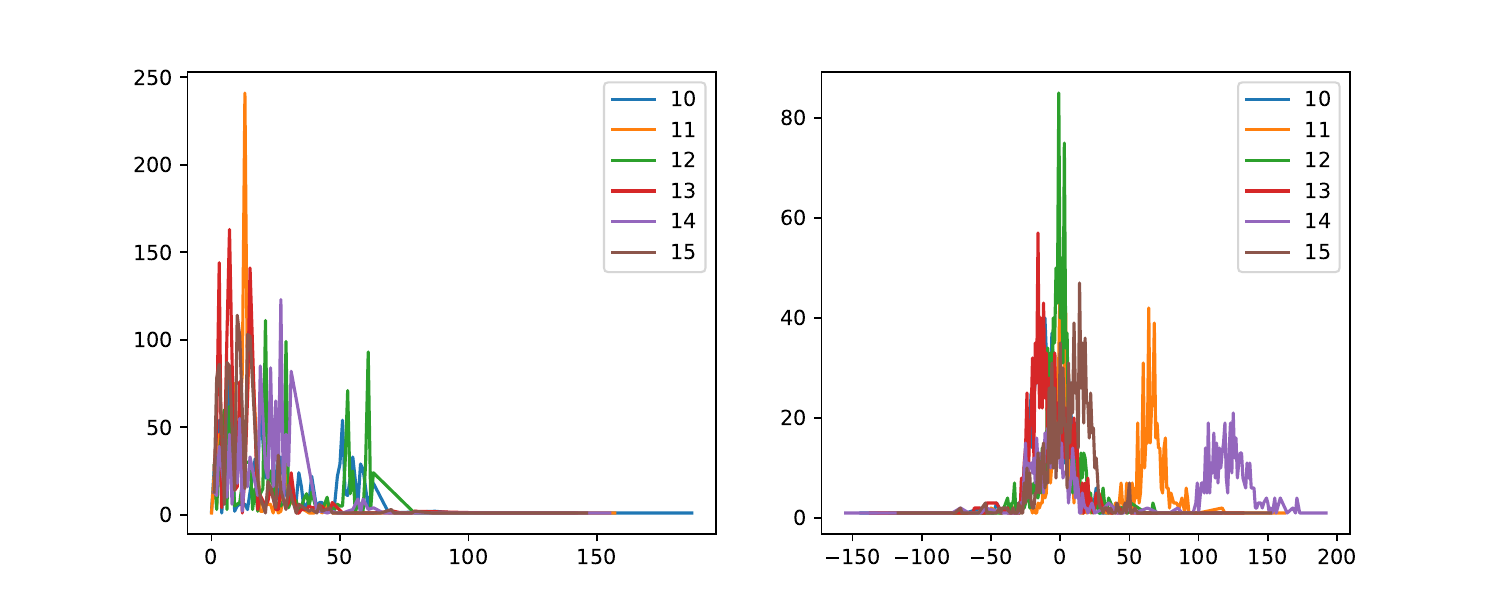}
    \caption{Example results for arc walk counter circuit (8 counter qubits, 10 to 15 steps).
On the left is a purely incrementing circuit, on the right is a two-way walk using the reversal and superposition technique above. 1000 shots for each number of steps.}
    \label{fig:arc_walk_counter_results}
\end{figure}

Example results (8 counter qubits, 10 to 15 steps) are shown in Figure \ref{fig:arc_walk_counter_results}. 
On the left is a purely incrementing circuit, on the right is a two-way walk using the reversal and superposition technique above.

\subsection{Random Jump Circuit}

In this class of examples, instead of using smaller rotations for higher-order qubits, we set up the circuit so that these qubits are changed less often. 

\begin{figure}
    \centering
    \includegraphics[width=\linewidth]{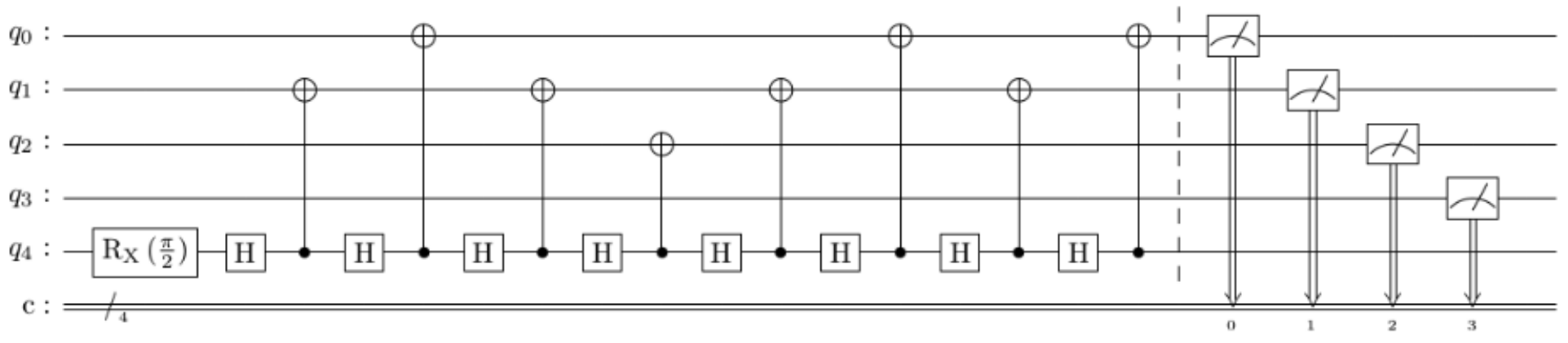}
    \caption{Random jump circuit}
    \label{fig:random_jump_circuit}
\end{figure}

This can be done with and without a quantum coin controlling the gates. The example in Figure \ref{fig:random_jump_circuit} uses a standard Hadamard quantum coin.

At each step, a different target qubit in the counter register is selected, according to some weighted random sampling function. This function should prefer the lower-order qubits). In the example below, the selection was done with the distribution $\{\frac{1}{2}, \frac{1}{4}, \frac{1}{8}, \ldots.\}$. Note that the circuit creation now introduces  classical randomness, as well as the circuit measurement still having quantum randomness.

\begin{figure}
    \centering
    \includegraphics[width=\linewidth]{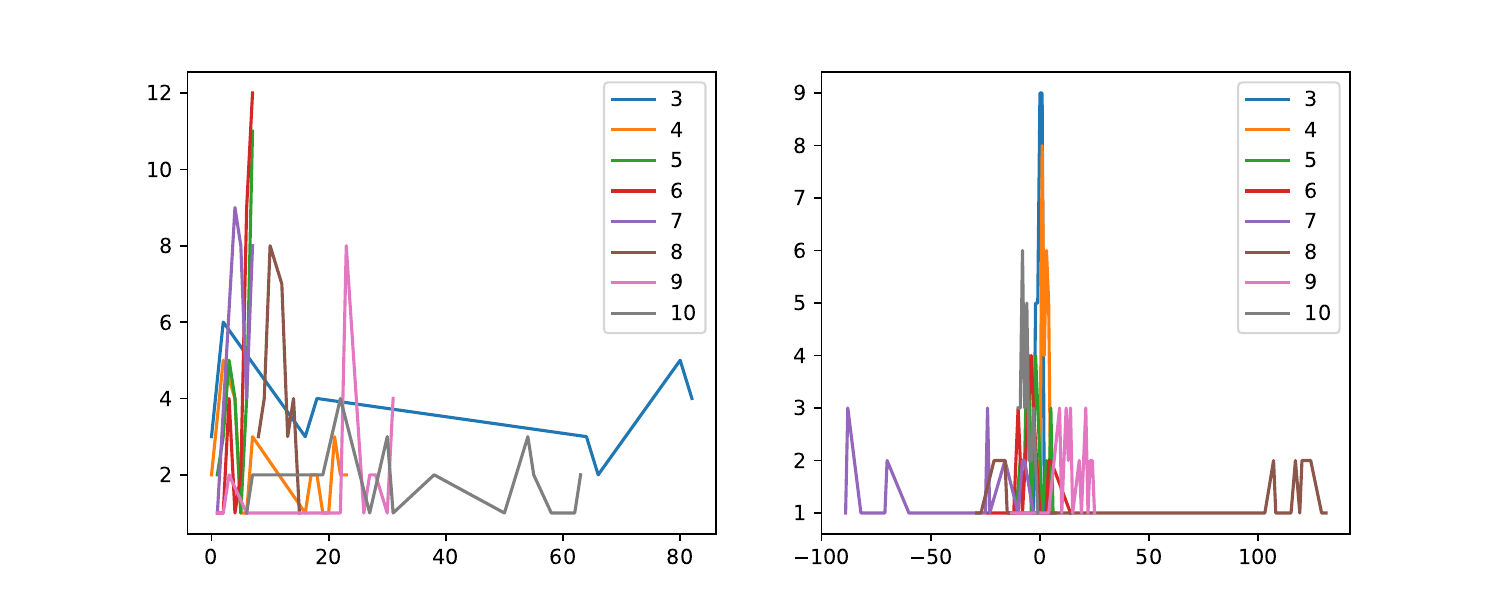}
    \caption{Random jump circuit results (positive only on the left,  two-way walk using reversal and superposition on the right). 30 random jump circuits per step, 30 shots per circuit.}
    \label{fig:random_jump_results}
\end{figure}

Results for up to 9 steps, using an 8 qubit counter register, are shown in Figure \ref{fig:random_jump_results}. As expected, each walk can go both ways (because randomly flipping a bit can reduce as well as increase a number), though the average tendency for an individual walk is to increase. This is because the registers are initialized to zero, so the process randomly diffuses out from zero.

\subsection{Cascading Disjunction Circuits}

\begin{figure}
    \centering
\includegraphics[width=6cm]{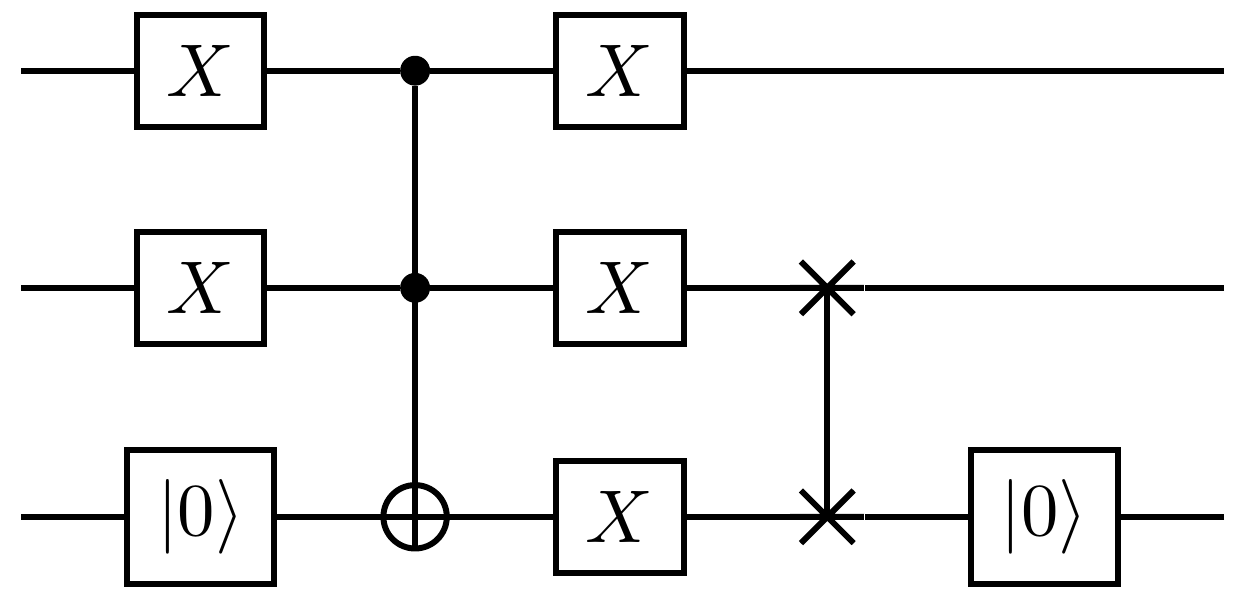}
    \caption{Boolean in-place disjunction circuit, which sets the state of the higher order
    qubit to the Boolean OR of the input states of the lower and higher order qubits. 
The $X$ gates surrounding the Toffoli gate implement the usual NOT operations to turn an AND conjunction operator into an OR disjunction operator $(A \vee B = \neg(\neg A \wedge \neg B))$. Finally, the swap gate and the reset to $\ket{0}$ operations ensure that the output is written back into the higher order qubit, and the ancilla qubit is reset to its initial state. Using this construction several times requires mid-circuit reset, so that the ancilla qubit can be reused.}
    \label{fig:boolean_or_inplace}
\end{figure}

`Cascading disjunction' is an extra circuit component that can be added to any of the counting circuits above. The idea uses a standard circuit component that performs Boolean disjunction, as in Figure \ref{fig:boolean_or_inplace}.

In this implementation, such gates are added between randomly chosen lower- and higher-order bits, using the same sampling distribution as in the Random Jump Circuits. This enables the values of lower-order qubits to propagate to higher-order qubits, which increases and tends to preserve progress in the walk, because the higher-order qubits are less likely to be randomly selected to be switched back to $\ket{0}$ later.

\todo{Pg 14, Figure 16: It is sometimes hard to interpret from the figures how much noise is due to the limited number of data points, and if the distribution changes much with the number of shots. This topic is discussed later but it might be useful to the non-specialist reader to flag it earlier.}
    
\begin{figure}
    \centering
    \includegraphics[width=\linewidth]{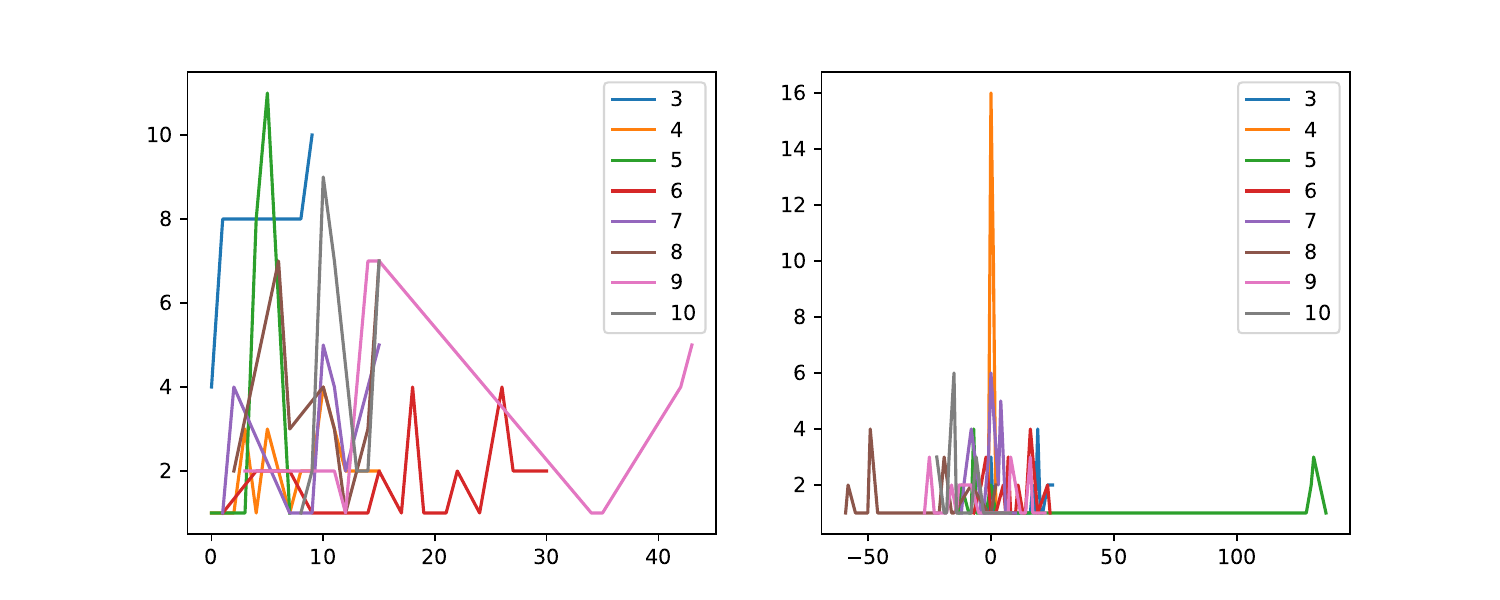}
    \caption{Cascading disjunction circuit results (positive only on the left,  two-way walk using reversal and superposition on the right). 30 random jump circuits with cascading disjunctions per step, 30 shots per circuit.}
    \label{fig:cascading_results}
\end{figure}

Results of including this technique are shown in Figure \ref{fig:cascading_results} (positive only on the left,  two-way walk using reversal and superposition on the right). The walks are still random, but propagating more reliably with the cascading disjunction components.

Another variant of this technique would be to add a conjunction between the coin qubit and the target qubit, that sets the next higher qubit to $\ket{1}$ before reversing the target qubit to $\ket{0}$. This would behave like a limited-carry operation: it performs some of the coordination between qubit values in the traditional bit-counter circuits in the literature, but much less, in a much more targeted fashion.

\section{Preliminary Results of Counting Circuits, With and Without Noise}

\def\hd#1{\parbox{1.3cm}{\centering #1}}

\begin{table}[t]
\small
    \centering
    \begin{tabular}{crrrrr}
\toprule
\hd{Steps} & \hd{Binary counter} & \hd{Arc counter} & \hd{Arc walk} & \hd{Random jump} 
& \hd{\footnotesize{Random jump with cascading} \scriptsize{disjunctions}} \\
\midrule
0& 0.000 & 0.000 & 0.000 & 0.000 & 0.000 \\
1& 1.000 & 1.108 & 0.491 & 4.386 & 1.133 \\
2& 2.000 & 3.276 & 1.522 & 3.383 & 5.281 \\
3& 3.000 & 4.742 & 2.079 & 8.488 & 5.783 \\
4& 4.000 & 6.764 & 2.826 & 11.450 & 6.341 \\
5& 5.000 & 8.233 & 3.833 & 9.317 & 13.539 \\
6& 6.000 & 10.564 & 4.860 & 12.468 & 15.510 \\
7& 7.000 & 10.955 & 6.476 & 13.964 & 16.366 \\
8& 8.000 & 12.664 & 6.458 & 9.804 & 21.479 \\ 
9& 9.000 & 15.097 & 6.958 & 14.261 & 16.837 \\
10& 10.000 & 17.836 & 7.810 & 12.261 & 21.919 \\
\bottomrule
\end{tabular}
\vspace{0.1in}
    \caption{Average distances traveled after $n$ steps, ideal simulation}
    \label{tab:results_ideal}

\vspace{0.2in}

    \begin{tabular}{crrrrrr}
\toprule
\hd{Steps} & \hd{Binary counter} & \hd{Binary counter (QPU)} & \hd{Arc counter} & \hd{Arc counter (QPU)} 
& \hd{Arc walk} & \hd{Random jump} \\
\midrule

0& 0.000 & 0.00 & 0.000 & 0.002 & 0.000 & 0.000 \\
1& 15.606 & 21.946 & 1.312 & 1.94 & 2.562 & 1.742 \\
2& 25.123 & 32.771 & 3.461 & 4.412 & 3.859 & 2.618 \\
3& 30.504 & 32.802 & 5.488 & 5.782 & 6.223 & 5.919 \\
4& 29.921 & 33.603 & 6.824 & 6.236 & 8.721 & 9.279 \\
5& 31.046 & 32.212 & 8.931 & 10.208 & 10.773 & 10.250 \\
6& 31.732 & 33.317 & 10.442 & 12.984 & 12.627 & 10.981 \\
7& 30.837 & 32.978 & 11.183 & 12.389 & 14.978 & 12.401 \\
8& 31.894 & 33.775 & 13.148 & 13.682 & 15.898 & 14.568 \\
9& 30.99 & 32.989 & 14.614 & 15.504 & 17.300 & 12.684 \\
10& 31.912 & 34.101 & 18.459 & 19.666 & 20.907 & 17.216 \\

\bottomrule
\end{tabular}
\vspace{0.1in}
    \caption{Average distances traveled after $n$ steps, noisy simulation and QPU}
    \label{tab:results_noisy}
\end{table}

The big advantage for the simpler models presented here is that we can run them much more accurately (and quickly) on NISQ-era quantum hardware.
Preliminary results for the different circuit designs presented in this paper
are given in Tables \ref{tab:results_ideal} and \ref{tab:results_noisy}.
The goal of the new circuits is more reliable representation and incrementing of position, so the key comparison is with a binary counter itself, rather than
the use of a binary counter going up and down in a random walk. The new methods
are evaluated just with $n$ positive steps, without subtracting the results of
a walk in the other direction. The use of cascading disjunctions is included in the
ideal `random jump' circuit, though not in the noisy simulation, because the support
for mid-circuit reset on the ancilla qubit is not guaranteed today.
The simulations used 6-qubit counting registers, so can represent numbers from 0 to 63.

Ideal simulations can only be performed for small numbers of qubits (as a rule-of-thumb, as we pass 30 qubits we start to break the
limits of classical simulation). 
In addition, the binary counter and arc counter circuits were run on a quantum computer
with 11 trapped-ion qubits, described by \citet{wright2019benchmarking}. These results
are also show in Table \ref{tab:results_noisy}, with the label QPU (quantum processing unit). 
Note that 11 qubits is relatively small 
by today's standards: the number of reliable qubits in state-of-the-art machines in 2023 
is at least in the 20s \citep{ionq2022aria}. The older machine was deliberately chosen for this 
experiment, because it highlights the frailty of exact binary counting circuits 
compared with approximation counting circuits.

The random jump results were computed using 30 shots on each of 30 randomly-generated 
circuits, so these results include both classical and quantum randomness. The binary
and arc circuits are generated deterministically, so these results include 1000 shots
for a single circuit, and all the randomness is quantum.

Key findings include:

\begin{itemize}
    \item The binary counter results are perfect with ideal simulation, but
    are rendered useless in the noisy simulation. They quickly converge to a random
    number around 32, which is the average 
    \item For all the other circuits, the difference between ideal and noisy results 
    is much less.
    \item The average results for the arc counter and arc walk circuits are 
    the most reliable for simulating a monotonically-increasing position, with or without noise.
    \item The random jump results also tend to increase, but tend to plateau and then
    move up and down randomly. (This randomness is smaller with a larger register.)
\end{itemize}

\begin{figure}
    \centering
    \includegraphics[width=0.7\linewidth]{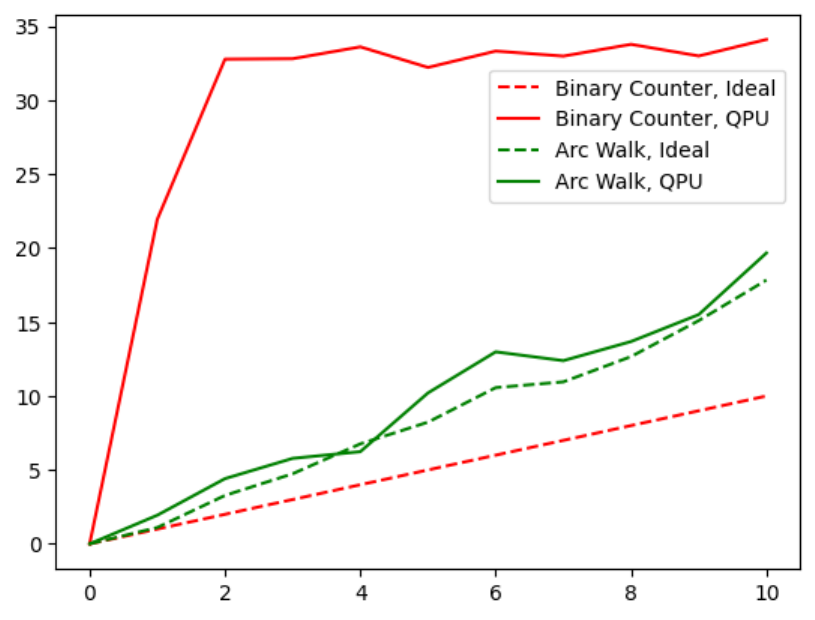}
    \caption{Ideal and actual QPU results for binary counter and arc walk circuits. The QPU are much closer to the ideal
    monotonically-increasing results for the arc counter, whereas they are useless after 2 steps for the binary counter}
    \label{fig:walks_on_qpu}
\end{figure}

The QPU results for binary and arc counting are compared graphically in Figure \ref{fig:walks_on_qpu}.
This shows how quickly the binary counter becomes useless on a real QPU, whereas by contrast, the arc counter 
QPU results stay close to the ideal simulated results.

\subsection{Quantum Walk Distributions and Real Financial Data}

The notion that market returns follow a normal (or log-normal) distribution is standard and established
in quantitative finance, even though it has been known for decades that heavy-tailed distributions
are sometimes a better fit \citep{mandelbrot1963economics,zi2017heavytailed}.
In the traditional random walk model, daily changes in asset price are also assumed to follow
a normal distribution, or even more simply, to take a constant step in either direction,
and the accumulation of many such steps in a binomial distribution eventually approximates the normal distribution.

\begin{figure}
\adjustbox{width=\linewidth}{
\def\figwidth{4cm}
    \centering
    \small
    \begin{tabular}{ccc}
    \includegraphics[width=\figwidth]{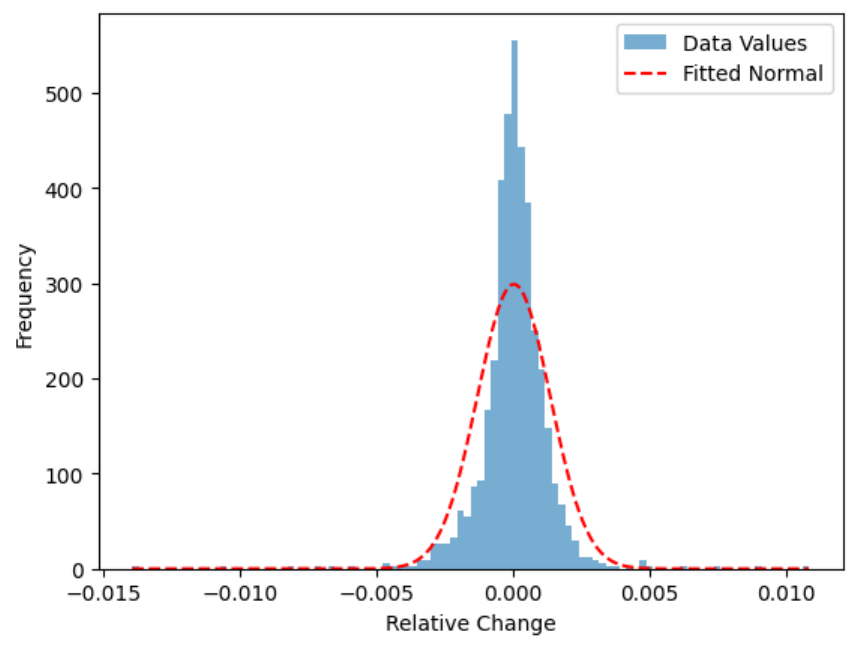} &
    \includegraphics[width=\figwidth]{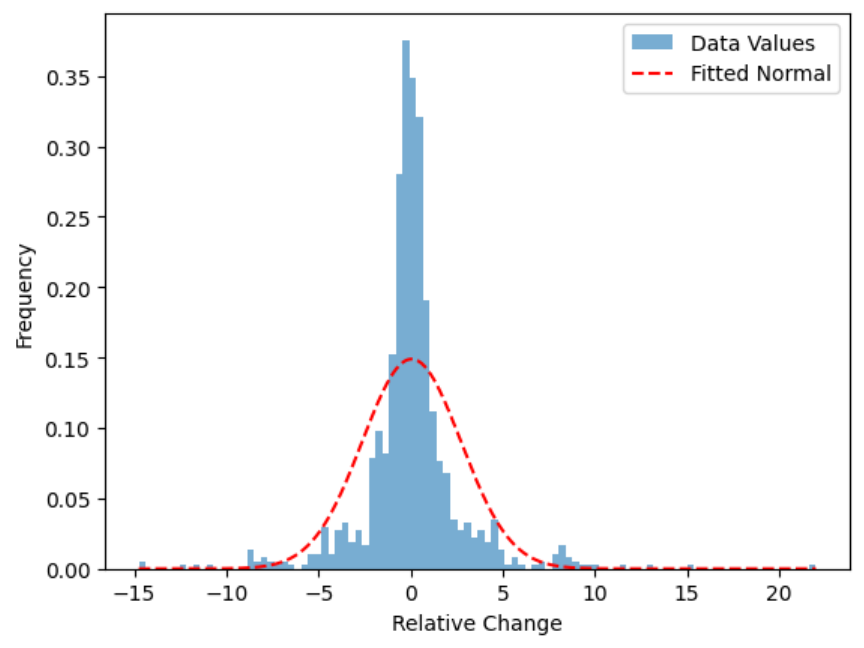} &
    \includegraphics[width=\figwidth]{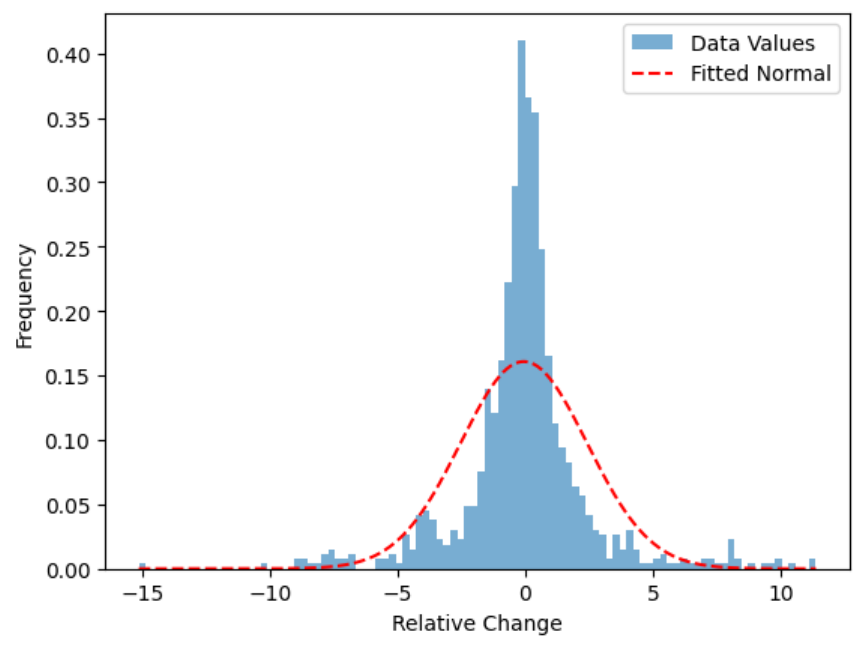} \\
    Dow Jones Index & Arc Walk Circuit & Random Jump Circuit \\
    5 yrs of data from & 8 qubits & 8 qubits \\
    Yahoo! Finance & 1000 shots per job & 8 circuits per sample \\ & & 1000 shots per job \\
    \end{tabular}
    }
    \caption{Distributions of relative changes in the Dow Jones Industrial Average, and in quantum approximate counting simulations. The dashed curve (red) shows the best-fit normal distribution, that is, the normal with the 
    same mean and variance.}
    \label{fig:djia-comparisons}
\end{figure}

In practice, daily relative changes in stock prices also tend to have a distribution where most values cluster around
zero, but significant outliers cause a normal distribution fitted with the same mean and standard deviation
to underestimate the density in the middle if the distribution. This is shown for the Dow Jones Industrial Average 
in Figure \ref{fig:djia-comparisons} (using data from the Yahoo! Finance API). 
An initial comparison shows that distributions of daily changes generated 
by the arc walk and random jump circuits also follow this heavy-tailed pattern, which is not modeled well by normal approximations.

This does not show that the quantum approximate counting circuits give a better prediction of stock price changes
on a specific day. But it does show that the distribution of possible changes can be better-adapted to real-world financial
data, without the artificial constraint that daily changes should be normally or uniformly distributed.

It should be noted that quantum results in Figure \ref{fig:djia-comparisons} are obtained with some
parametrization and averaging, because all quantum job results are averaged over the number of shots,
and the random jump circuits are averaged over a number of sample circuits as well. The impact of long-tail
measurements depends on how large a sample we take. While this implies that there is some arbitrariness in results,
it also means that parameters such as the number of qubits, circuits, and shots can be tuned to model particular datasets.

In related work, IonQ quantum computers have also been used to model the normal distribution itself, using
a matrix product state technique that can readily be adapted to other distributions, because it relies on piecewise polynomial 
approximation \citep{iaconis2023quantum}. 
One of the longer-term promises of such work is that such distributions can be used as inputs for models such as the Monte Carlo simulations
advocated by \citet{egger2020quantumfinance}. If we have a reliable circuit for preparing a particular distribution, then such a
circuit could be used as input for Monte Carlo modeling by entangling its output with the simulated variables, rather than by 
sampling an individual number from the distribution and using this as a single `classical' random input value. 

\section{Quantum Walks with Mid-Measurement: A Quantum Zeno Effect}
\label{sec:mid-measure}

A crucial difference between classical random walks and quantum walks is that quantum walks behave differently depending
on when they are measured. There is no classical counterpart for this behavior, because a hallmark of classical systems 
is that their state is revealed but not changed by measurement.

In theory, it is possible to prevent a quantum system from changing state at all, by measuring smaller and smaller intervals.
As a simplest example, the gate $R_X(\theta)$ from Figure \ref{fig:single-qubit-gates} operating on a qubit in state $\ket{0}$ produces the state
\[ R_X(2\theta)(\ket{0}) = \cos(\theta)\ket{0} + i\sin(\theta)\ket{1}. \]
The probability of transitioning to the state $\ket{1}$ is thus proportional to $\sin^2(\theta)$ which tends to zero for small $\theta$, and it is easy to see that if a larger angle is divided into smaller and smaller increments, the probability
of observing a transition to the state $\ket{1}$ in {\it any} of these increments also tends to zero, because 
$\lim_{n\rightarrow\infty}(n) \sin^2{n} \rightarrow 0$. 

This phenomenon is sometimes called the quantum Zeno effect, after Zeno's classical paradox of motion. 
Of crucial interest for this paper, such effects have also been observed in psychology.  
\cite{kvam2015interference} demonstrated that participants are likely to form less extreme judgments of
moving scenes if asked to judge the motion in smaller time-frames, and \cite{yearsley2016zeno} demonstrated
that participants evaluating evidence in a criminal trial are more likely to change their minds if 
several pieces of evidence are presented before asking for a decision.

\begin{figure}
    \centering
    \includegraphics[width=8cm]{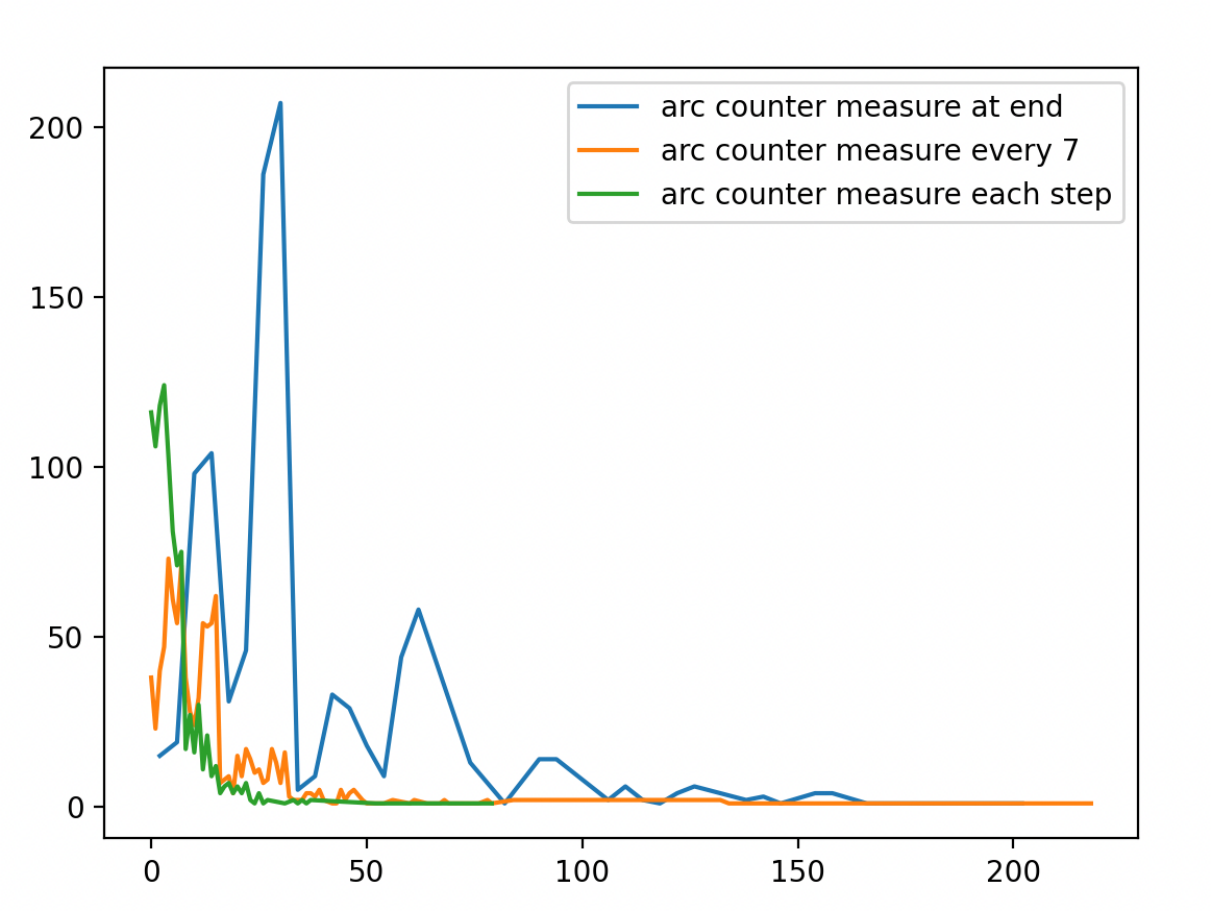}
    \caption{Arc counter circuit results, simulating the results of quantum walks with 20 steps, with and without mid-circuit measurement at the given positions.}
    \label{fig:arc_counter_mid_meassure}
\end{figure}

It is easy to add mid-circuit measurement to our quantum approximate counting circuits and to evaluate the results,
at least in simulation. (The availability of mid-circuit measurement varies across quantum platforms currently, 
partly because the accuracy of the measurement and reset operations is hard to guarantee.) Example results are shown in Figure \ref{fig:arc_counter_mid_meassure}, simulating walks with 20 steps, with no mid-measurement, measurement every 7 steps, and measurement every step. The average positions reached by these walks were 35.6, 14.3, and 5.7 respectively, so as expected,
the use of mid-measurement reduces the average distance traveled in the quantum walk. (It is not always this simple, particularly due to periodicity.) \revision{A further step for this research would be to experiment with parameters
including the number of steps, distribution of step sizes, and frequency of measurements, to see if different levels of
deliberate decoherence can bring the quantum results of Figure \ref{fig:arc_counter_mid_meassure} closer to the distributions
observed with the Dow Jones Index and other asset prices.}
    \todo{Pg 18, Figure 17. Could the author discuss whether it is possible to add a tuneable level of noise/decoherence to the quantum circuits (see also earlier comment about oscillators), so that the quantum results could be brought closer to the DJIA index results?}

\subsection{Mid-Measurement, Transactions, and the Housing Market}

In the quantum economics theory of \citet{orrell2020quantumeconomics}, the act of measurement is compared with
fixing a transaction, and subjective opinions of value can vary like quantum states between transactions.
By analogy with the quantum Zeno effect, we may expect that opinions about 
prices vary less if there are more frequent measurements, i.e. more frequent transactions. 
Evidence is presented by \citet{orrell2022oscillator,orrell2023keep} 
showing that price volatility is not constant, and that high volatility corresponds to both uncertainty in value,
and wide ranges between different bid prices and asking prices. 
\revision{This section presents some findings from housing market data, indicating that indeed, larger transaction 
volumes correlate with smaller differences between the higher prices asked by sellers, and lower prices offered by buyers.} 

\todo{The idea of nearby transactions corresponding to a measurement is very intriguing. But, I could not quite get the point about low vs high sell prices. It definitely seems that the author is on to something, but, at the very least, I would suggest revisiting the corresponding text with an eye for clarity. Note, in these parts of the paper (which for some readers might be the most interesting), it would pay off to be more clear about the source of quantum-like advantage.}

The range of prices offered to buy or sell an item is typically much more apparent in the housing market than the stock market, 
because each item for sale is priced and negotiated much more individually. A standard process involves the use of
comparable sales or comps, in which transactions on properties nearby in time, space, and value are used to form a pricing estimate \citep{pagourtzi2003real}. If these nearby transactions correspond to measurements of the system, then the quantum Zeno
effect would suggest that the outcome of this measurement is more certain if there are more nearby transactions.

This effect was demonstrated in practice using the following modeling assumptions, and summary data published by Zillow.

When a house sells for less than its original listing price, we assume that 
this indicates a difference between the seller’s and the buyer’s estimate of the house’s value. Larger uncertainties in the 
market would support larger differences of opinion. Even when considering monthly averages of data, we would expect that a 
smaller number of sales in a given area would contribute to greater market uncertainty, and this should correlate with a greater 
difference between the list price and the sale price. 

By contrast, when a house sells for {\it more} than its original listing price, we assume that there are other 
factors involved: in particular, this situation is common when there are other bids on the property from 
other potential buyers, so a minimum 
value is already established without the need for previous comparable transactions. 

Thus, we assume that the markets where lack of comparables is a primary factor in price uncertainty 
are those where the average sale price is less than the average list price. 
\revision{It follows that, if we restrict our attention to markets where the average sale price is less 
than the average list price, we should see evidence that lower transaction volumes are correlated with 
greater disparities between list price and sale price.}

Data used to test these hypotheses was gathered from the Zillow Housing Data portal\footnote{\url{https://www.zillow.com/research/data/}, accessed 2023-10-05.}.
The datasets are summary statistics: counts and averages. These are only comparable within a given metro area: for example, 2000 sales in a month would be very low for New York, NY, and very high for Wichita, KS. Thus we compute correlations by comparing monthly statistics within each metro area.

The algorithmic steps are as follows:

\begin{itemize}
    \item For each metro area:
    \item For each month:
    \begin{itemize}
\item Collect the sales count and the average list-to-sale price ratio.
\item If the average list-to-sale price ratio is greater than 1, skip this month.
    \end{itemize}
\item This gives a set of (count, ratio) pairs, e.g. [(822, 0.98), (785, 0.96), (803, 0.97)], etc.
\item Compute the Pearson correlation coefficient between these sales counts and list-to-sale price ratios.
\item Gather the Pearson correlation coefficients into a histogram to see if there is a general trend.
\end{itemize}

\begin{figure}
    \centering
    \includegraphics[width=0.6\linewidth]{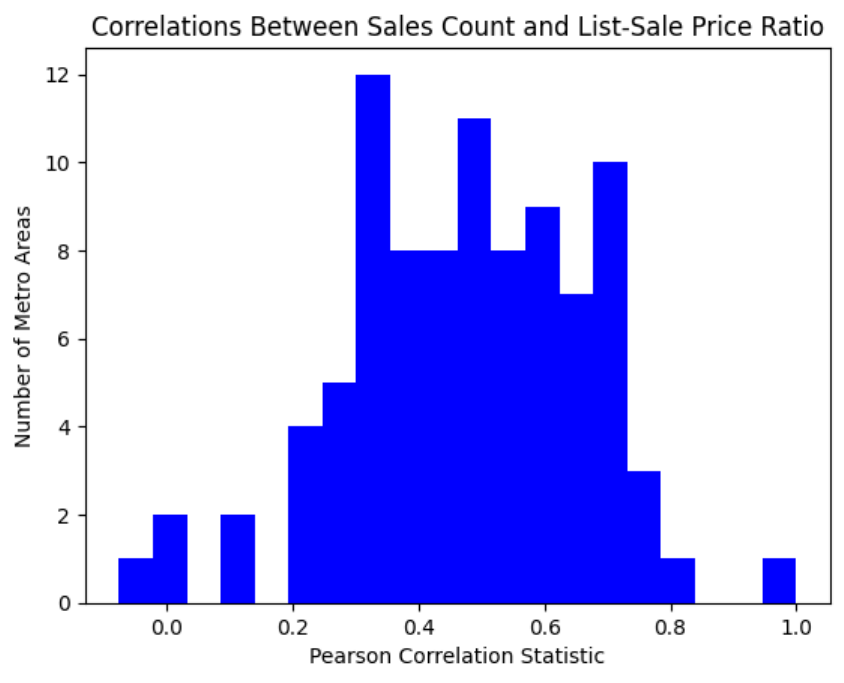}
    \caption{Histogram showing correlations between larger numbers of transactions and smaller list-to-sale price differences. Data from Zillow Housing Data.}
    \label{fig:housing_metro_correlations}
\end{figure}

The result is in Figure \ref{fig:housing_metro_correlations}. Nearly all the correlations are strongly positive. This shows that, in cases where a house is sold for less than its asking price, there is a very strong correlation between the translation volume, and the closeness of the list and the sale prices.

\begin{figure}
    \centering
    \includegraphics[width=0.8\linewidth]{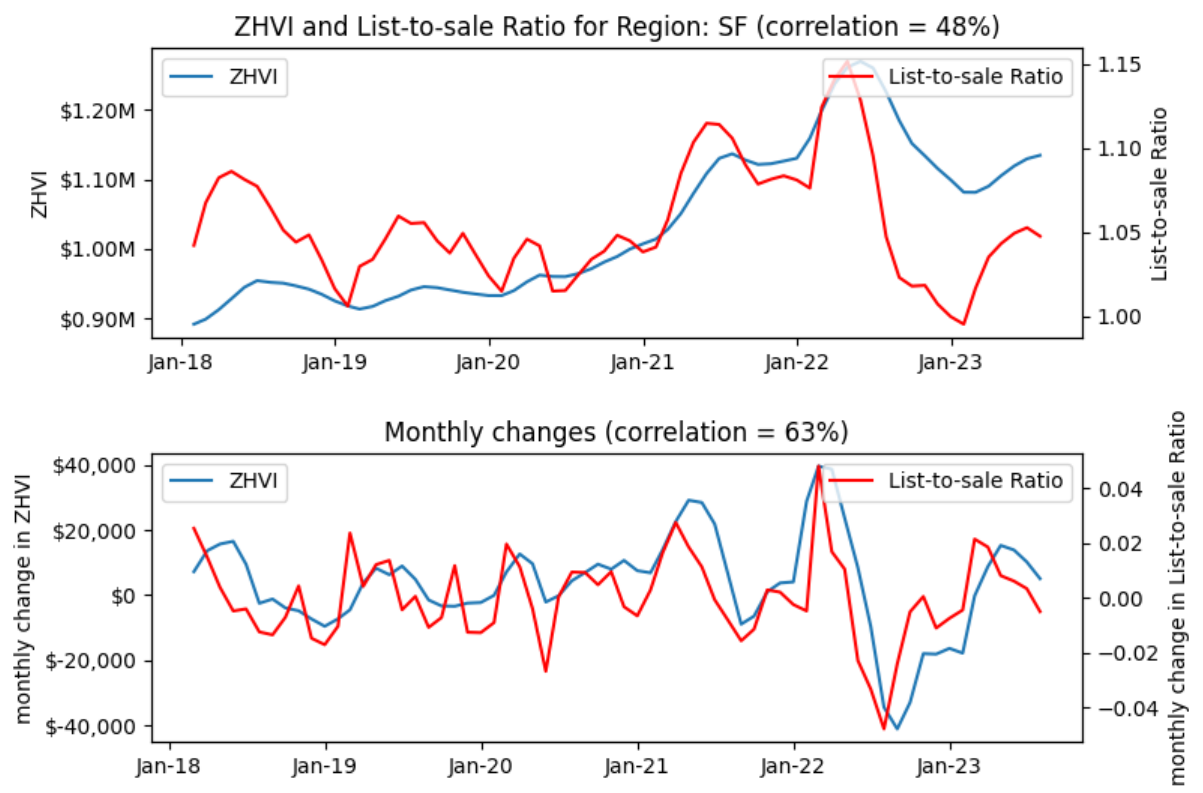}
    \caption{Monthly values of Zillow Home Value Index (ZHVI) and list-to-sale ratio as well as monthly changes that show positive correlation in both rising and falling markets.}
    \label{fig:housing_metro_time_series}
   
\end{figure}

    \todo{Pg 21, Figure 19. This might be beyond the scope of this paper but it would be interesting to see how the price behaviour changes for the case where prices are rising vs falling.}

\revision{
In addition to the number of sales in a region, it is instructive to look at the correlation between home prices and the list-to-sale price ratio since the price of a home is the driving factor in a home purchase.  As a proxy for home prices we use the Zillow Home Value Index (ZHVI) which reflects the typical value for homes in the 35th to 65th percentile range.

Figure \ref{fig:housing_metro_time_series}
shows both the raw monthly values (above) for the San Francisco region as well as  monthly changes (below) comparing the ZHVI and list-to-sale ratio values for the last 6 years.  The bottom chart shows more clearly the positive correlation between the two if we consider monthly changes.  Monthly changes give a better sense of market ups and downs.  As prices fall, we see that buyers are more likely to bid lower than the list price.  Conversely, in a rising market, buyers are more likely to bid more than the list price.
}

That fewer transactions correlates with larger list-to-sale price ratios is in line with the trends 
expected from quantum economics models, in which various beliefs and opinions about value can evolve 
and diverge more when there are fewer transactions or measurements. 
\revision{The correspondence between overall price changes and list-to-sale ratios could be
described by a quantum walk with a clear directional momentum.}
However, it is also easy to propose simple non-quantum models 
for these behaviors. Fewer comparable samples leads to greater sampling error and thus greater price uncertainty:
thus lower liquidity brings higher volatility.
One potential
strategy for evaluating and distinguishing which approaches are better would be to consider the dynamics / evolution of prices
in such models: for example, to try applying the quantum Monte Carlo sampling reduction described by \citet{egger2020quantumfinance} to the problem of making accurate price estimates with fewer comparable sales.

\subsection{Macro Effects: Collective Beliefs and Reckonings}

\todo{However, a key challenge in modelling eg pricing data is that measurement and the corresponding constructive influence would somehow need to have a macro interpretation.}

\revision{As well as the belief states of individuals,
quantum models have been proposed to represent the beliefs and ideologies of whole groups of people
\citep{kitto2016ideologies}.
With pricing models, the influence of contrasting beliefs held by groups of people is especially important
--- as \citet{bachelier1900theorie} wrote, 
``Contradictory opinions in regard to these fluctuations are so divided that at the same instant buyers believe the market is rising and sellers that it is falling.''
The importance of group beliefs is particularly apparent with rare luxury goods: the opinion of a few wealthy people 
can be enough to establish high prices for rare art, even without mass appeal. 

In such a model, decision making becomes partly a constructive process.
This is standard in quantum models for individual behavior, where participants 
tend to remain in a particular state once it has been chosen if an identical 
question is asked \citep{busemeyer2012quantummodels}.
Memory and recollective processes are studied from a quantum point of view by
\cite{waddup2023temporal}, who try to measure the influence of intervening steps in decision-making. 
One problem in demonstrating quantum properties, such as violations of temporal Bell inequalities, in 
these processes is that it is hard to rule out interference and signaling.

With financial decisions, `signaling' is a clear part of making decisions collective. Historic examples
are notable in the early telegraph era, where the ability to publish outcomes of horse-races and 
stock-transactions across large distances led to opportunities for 
short-term fraud and long-term financial services industries \citep{standage1998victorian}. 
Applying the notion of quantum measurement and state collapse to a whole group of people may sound
far-fetched in the abstract, but less so when we consider how much modern infrastructure is built to
make a decision in one place felt immediately everywhere. More generally, there are many social
processes where the views of groups of people are changed, sometimes reluctantly,  
in the face of events --- here the notion of a `transaction' becomes the even more general `reckoning'.

\todo{
Pg 22, line 33: “Any quantum ‘modeling advantage’ on this problem would be especially compelling.” On the other hand speed isn’t much of an issue when analyzing the real estate market, so it could be pointed out that the underlying sales dynamic will affect other asset types such as stocks and options where speed is more important. Also the quantum advantage involves not just computational speed, but also different modelling techniques which are based on quantum logic (even if they can be emulated classically).
}

Such collectively-recognized events are naturally scarce: we cannot create housing transaction examples in
the same way that we can collect and annotate training examples with other supervised machine
learning models.
Any quantum `modeling advantage' with such problems would be especially compelling, because when
training data is scarce, we cannot just train larger classical models for longer and assume they will give better results.
Quantum Monte Carlo pricing models are expected to converge faster with fewer samples \citep{rebentrost2018quantum},
which raises the intriguing question of whether such techniques lead to faster convergence with fewer examples from
real experience. Even with other asset types such as stocks and options where transaction samples are plentiful and 
speed is more important, the dynamic effects of large events from other market sectors could be anticipated 
more effectively. There is still scope for the use of quantum models in these areas,
even before the widespread use of quantum computers.
}

\section{Conclusions and Future Work}

This paper has introduced and explored quantum approximate counting circuits, as fault-tolerant alternatives to the traditional quantum walk design, particularly for the way position is tracked and incremented.
The new designs presented here lack some of the mathematical elegance, and the theoretical results, that accompany the traditional quantum walk design: and in particular, there are no longer unit increment and decrement operators that correspond to the ladder operators of a quantum oscillator. 
However, the enormous advantage for the simpler models presented here is that 
they behave much more accurately on NISQ-era quantum hardware, which could contribute to commercially advantageous 
applications of quantum computers in economics.

These are just prototype designs so far. The main next steps for this work are to evaluate the proposals
more quantitatively, answering the following two questions:

\begin{enumerate}
    \item How do results on NISQ-era quantum computers correspond to ideal or simulated results for small
    circuits, and what does this indicate about the expected behavior on quantum hardware for systems that are too big to simulate on classical hardware?
    \item How do results compare with the distributions observed with real market behaviors?
\end{enumerate}

The ideal outcome of this research is that we would find circuit walk designs that are robust enough
to given better models of market behavior that include some of the benefits of quantum approaches
noted by \cite{orrell2020quantumeconomics}, while being able to run on today's quantum hardware without waiting for error-correction.

Given the crucial and explicit role that measurement plays in quantum models, it is possible that some of the earliest
such quantum advantages will be apparent in markets where a small number of significant transactions can dramatically
influence the price of a particular asset. An initial analysis suggests that the housing market may be an appropriate
area to test this hypothesis.

This work can be seen as part of a larger program to bring value in economic modeling on quantum computers.
Other successes for quantum circuit designs include modeling and sampling from key distributions \citep{iaconis2023quantum}, 
and demonstrating particularly effective time-series models using copula functions implemented using entanglement \citep{zhu2022copula}. Related work in cognitive science has demonstrated that simple quantum circuits
can also be used to model decision-making processes \citep{widdows2022cognitivecircuits}. In the next few years, it is likely that several
such small components, being developed today, will be used as key building blocks in the first 
profitable applications of quantum computing in economics.

\section{Acknowledgements}

The author would like to thank Emmanuel Pothos and David Orrell for interesting conversations and encouragement.

\section{Funding}

This work was funded by IonQ, Inc.

\ifarxiv
    \small
    \bibliographystyle{SageH}
    \bibliography{ionq}
\else

\fi

\end{document}